\documentclass[11pt, oneside]{amsart}   	
\usepackage{geometry}                		
\usepackage{graphicx}				
\usepackage{amssymb}
\usepackage{geometry} 
\usepackage{amsmath,amsthm,amsfonts}
\usepackage{amssymb}
\usepackage{latexsym}
\usepackage[applemac]{inputenc}
\usepackage{tensor}
\usepackage{graphicx}
\usepackage{bm} 
\usepackage{mathabx}
\usepackage{pdfsync}
\usepackage{fancyhdr}
\usepackage{float}

\usepackage[T1]{fontenc}
\usepackage{color}
\usepackage{xcolor}
\usepackage[breaklinks,colorlinks,backref]{hyperref}
\hypersetup{
    colorlinks, 
    linktoc=all, 
    linkcolor=red,  
  citecolor=hyptxt,
  urlcolor=blue}
  \hypersetup{
  citebordercolor=,
  filebordercolor=green,
  linkbordercolor=blue
}
\definecolor{hyptxt}{rgb}{0.7, 0.4, 0.9}
\usepackage{bibtopic}

\definecolor{hervecolor}{rgb}{0.8,0,0.7}
     
\newcommand{\ket}[1]{|\kern.3ex#1\kern.3ex\rangle}
\newcommand{\bra}[1]{\langle\kern.3ex #1 \kern.3ex|}
\newcommand{\scalar}[2]{\langle\kern.3ex #1 \kern.3ex|\kern.3ex#2\kern.3ex\rangle}

\newcommand{\ii}{\mathsf{i}}

\def\R{\mathbb{R}}

\def\ud{\mathrm{d}}

\def\mfn{\mathfrak{n}}

\def\bsb{\boldsymbol{\beta}}

\numberwithin{equation}{section}

\begin{document}

\title[Quantum Mixmaster as  a model of the Primordial Universe]{Quantum Mixmaster as a model of the Primordial Universe}
\author{Herv\'e Bergeron}
\address{ISMO, UMR 8214 CNRS, Univ Paris-Sud,  France}
\email{herve.bergeron@u-psud.fr}

\author{Ewa Czuchry}
\address{National Centre for Nuclear Research, 00-681
Warszawa, Poland}
\email{Ewa.Czuchry@ncbj.gov.pl}

\author{Jean-Pierre Gazeau}
\address{APC, UMR 7164 CNRS, Univ Paris  Diderot, Sorbonne Paris Cit\'e, 75205 Paris, France}
\email{gazeau@apc.in2p3.fr}

\author{Przemys\l aw Ma\l kiewicz}
\address{National Centre for Nuclear Research, 00-681
Warszawa, Poland}
\email{Przemyslaw.Malkiewicz@ncbj.gov.pl}


\begin{abstract} The  Mixmaster solution to Einstein field equations was examined by C. Misner in an effort to better understand the dynamics of the early universe.  We highlight the importance of the quantum version of this model for early universe. This quantum version and its semi-classical portraits are yielded through affine and standard coherent state quantizations and more generally affine and Weyl-Heisenberg covariant integral quantizations. The adiabatic and vibronic approximations widely used in molecular physics can be employed to qualitatively study the dynamics of the model on both  quantum and semi-classical levels. Moreover, the semi-classical approach with the exact anisotropy potential can be effective in numerical integration of some solutions. Some promising physical features such as the singularity resolution, smooth bouncing, the excitation of anisotropic oscillations and a substantial amount of post-bounce inflation as the backreaction to the latter are pointed out. Finally, a realistic cosmological scenario based on the quantum mixmaster model, which includes the formation and evolution of local structures is outlined.
\end{abstract}

\maketitle
\tableofcontents


\section{Introduction}
\label{intro}

In the present contribution we aim to give an overview of our recent results on the quantum mixmaster universe while emphasising its potential relevance for modelling the primordial universe. We contrast the idea of the quantum mixmaster as a model of the primordial universe with some other existing theories.

\subsection{Motivation}
Observational data suggest that the observable Universe has emerged from its primordial phase in a very peculiar state: as a patch of flat, isotropic and homogeneous space furnished with tiny adiabatic density perturbations with amplitude that is nearly scale-invariant \cite{planck15}. Although it is not yet experimentally confirmed,  the Universe is believed to be also filled with gravitational waves that must have originated about the same time and since then must have propagated across space almost freely. The primordial gravitational waves (PGWs) are expected to extend across a wide range of wavelengths and therefore to provide an excellent probe of the primordial universe \cite{boyle06}. Although nowadays only upper bounds on the amplitude of PGWs are known, they may be detected soon as promising experiments are being developed \cite{tamago,litebird,core}.

Presently, the most developed theory of the primordial universe is the theory of cosmic inflation. It explains the origin of the primordial structure with a simple mechanism of the amplification of quantum vacuum fluctuations that seem to naturally inhabit the primordial space at sufficiently small scales. Despite its widely acknowledged success, inflation has two inherent drawbacks: it {\it postulates} the existence of an unknown field, the inflaton, in a fine-tuned potential, which dominates the primordial universe and the inflationary spacetime is geodesic past-incomplete. The latter makes the model sensitive to the unspecifiable initial condition. Furthermore, in the aftermath of the Planck mission, the inflationary paradigm seems to have lost some of its original appeal \cite{listeinloe13} as the well-known problems of inflation such as the initial condition problem, the fine-tuning of the potential and the multiverse problem (or the  ``unpredictability'' problem) are now intensified. Needless to say, invoking the anthropic principle is necessary for this theory to work.

Alternative theories to inflation often connect models of singularity resolution with the origin of structure in the Universe \cite{khoury_etal01,petpint08}. They show that contraction preceding a quantum bounce can generate the initial density perturbations via an amplification mechanism analogous to the one which operates during inflation. Moreover, alternative theories generically predict the power spectrum of PGWs that is completely different from the inflationary one, which makes their case very attractive for experimenters. However, alternative models have difficulty in matching the exact CMB data as their predicted density perturbation spectrum tends to be blue-tilted, contrary to the best fit from the Planck data which is sightly red-tilted. To remedy this problem they often introduce ad hoc ingredients such as exotic fluids to drive the contraction that had preceded the bounce in a presupposed way. However, to our minds, the {\it real} problem with alternative theories is that they are based on the perturbed Robertson-Walker metric that is too simplistic to reasonably describe the bouncing dynamics and needs to be replaced with a more generic one.

\subsection{Our approach}
However speculative, inflation is still the leading theory of the origin of primordial structures after almost four decades from its inception  and despite the fact that a few alternative theories have been proposed to challenge the inflationary paradigm. Thanks to the ever increasing quality of the observational data, all new theories of the origin of primordial structure have to pass increasingly stringent tests. At the same time, new theoretical proposals should be at least in some respects essentially different from the available ones. 

 Our approach starts with a few assumptions. First, the Universe was dominated by quantum gravity effects in its primordial phase, which led the Universe to avoid the initial singularity through a bounce. Second, this very important cosmological event was at the heart of a physical mechanism that has generated the primordial structures and started the present cosmological expansion. Third, any restrictive a priori assumption on the primordial matter or the primordial symmetries should be avoided as much as possible. 
  
 In particular the last assumption can be implemented by removing isotropy of the background and employing the generic spatially homogeneous model, the Bianchi Type IX  (or  simply,  mixmaster) \cite{misner69}. And indeed, our study of the quantum and semi-classical mixmaster dynamics reveals that removing isotropy is actually necessary for a description of the primordial universe as the isotropy breaks down at the quantum bounce. The effect of anisotropy turns out to be a key feature of the quantum mixmaster dynamics leading to a complex and multi-stage quantum bounce. The latter includes anisotropic oscillations, an extended phase of accelerated post-bounce expansion and the coupling between different modes of perturbations.  This makes quantum mixmaster a very promising model of the primordial universe in which local structures are formed and evolve. In particular, this model may give rise to a novel and effective mechanism for generating primordial perturbations without postulating extra degrees of freedom and fine-tuned potentials.

 Since we propose to describe the primordial universe with a quantum model, we must face the issue of quantization and, in particular, of quantization ambiguity. A noteworthy aspect of our approach is the use of integral covariant quantization methods. They provide us with a rich world of quantum models which are valid for a given classical model. Thereby we equip the primordial universe model with a reasonable freedom by means of unspecified parameters. The possibility to adjust these parameters to observational data once the model is sufficiently developed is an integral feature of our approach. Furthermore, our quantization procedures allow to regularise  classical  singularities, contrarily to the commonly implemented canonical  quantization. Hence, we are able to select a quantization that smooths irregularities present in the classical model ((as is the case with the use of the affine and Weyl-Heisenberg covariant integral quantizations - see below for more details) and thereby simplify the subsequent quantum analysis without loss of relevant quantum features.

\subsection{Former results on mixmaster}
Let us briefly recall some basic facts about the Bianchi Type IX model of general relativity. This model was first investigated by Bogoyavlensky (see \cite{bogoyavlensky85} and references therein). Later on, it was found by Belinskii, Khalatnikov and Lifshitz  (BKL) that when an inhomogeneous universe approaches the singularity, time-like derivatives start to dominate the space-like and the universe enters an ultra-local phase of dynamics in which each spatial point evolves in accordance with the homogenous models' dynamics. Therefore, the dynamics of the mixmaster universe is considered as a prototype of this near-singularity generic behvior \cite{bekhali70,belkhali82}. Moreover, as a feasible fully non-perturbative approach to quantum gravity is not available, the mixmaster dynamics is perhaps the most complex cosmologically relevant model that can be tackled non-perturbatively.

The Hamiltonian formalism of the mixmaster dynamics was derived by Misner \cite{misner69}. His canonical formalism describes the universe in terms of a particle moving in the $3$-dimensional Minkowski spacetime in a time-dependent potential (note that this 3-d formalism describes the physical dynamics of a family of 4-d spacetimes). Misner has proposed a quantization of the system, however, it turned out that his quantum theory did not resolve the singularity. Since then no substantial progress has taken place on the quantum level. 

The {\it classical} mixmaster universe is known for its rich dynamics on approach to the singularity. As it contracts it undergoes anisotropic deformations of space, which in the vicinity of the singularity develop into chaotic oscillations. They can be viewed as two coupled polarization modes of a long nonlinear gravitational wave which is coupled to, and oscillates in, the isotropic universe. The energy of the wave grows rapidly and eventually takes over the dynamics. It has been  recently proven that a reasonable {\it quantization} procedure replaces the classical mixmaster singularity with a nontrivial quantum bounce \cite{berczgamapie15A,berczgamapie15B,berczgama16A,berczgama16B}. 
It was shown that due to the interplay between isotropic and anisotropic degrees of freedom the quantum bounce almost always involves complex dynamics. In particular, the anisotropic oscillations are generally produced at the quantum bounce in sufficiently large quantities to cause an extended phase of accelerated expansion to occur immediately after the bounce. In this light, the quantum mixmaster model appears to be a rich and promising model of the primordial universe, whereas quantum models of the Friedmann universe with a simple, symmetric bounce seem too simplistic. Actually, it was found that the Friedmann bounces correspond to very peculiar, {\it adiabatic} solutions in which the nonlinear wave is not excited  at all and therefore has little influence on the dynamics of the universe. 

The theory of perturbations around the {\it classical} mixmaster universe was studied by Hu, Regge, Parker, Fulling and Slagter \cite{huregge72,huA73,huA74,huB74,hufupar73,slagter83,slagter84}. Hu, Fulling and Parker investigated the quantum dynamics of the perturbations on the classical mixmaster in \cite{hufupar73}, where among other things they discussed the new phenomenon of mode-mixing between different spherical modes in the anisotropic universe, which is described by the so-called {\it general} Bogoliubov transformation. Hu has numerically solved the equations of motion for several modes of the scalar perturbation \cite{huA73,huA74}. Slagter has investigated the evolution of high-frequency gravitational waves \cite{slagter83,slagter84}. Although those results can give useful insights into the mixmaster perturbation theory, they cannot be straightforwardly applied within the framework of our approach. Firstly, they only partially develop the mixmaster perturbation theory and secondly, our method of quantization rests on the Hamiltonian formalism, which has not been really developed hitherto.

\subsection{Outline of the article}

The present article is mainly devoted to review our recent results on the quantum mixmaster. We focus on the questions of quantization of the classical model and of the subsequent semi-classical aspects, and on our treatment of the corresponding quantum and semi-classical dynamics through various approximations, some of them being similar to those in use in other domains, especially in quantum molecular physics. 

Section  \ref{classmod} provides the most essential definitions of the Bianchi Type IX model in its Hamiltonian formulation.  
In Section  \ref{quantmod} we give a general overview of the covariant integral quantization when the groups underlying the covariance are the affine group and the Weyl-Heisenberg group, the former standing for the symmetry of the open half-plane whilst the latter for the plane. These two symmetries are present in the mixmaster model. The first one concerns the scale factor, or volume, together with its conjugate momentum, forming the half-plane phase space $\sim \R^+ \times  \R$ for the isotropic part of the geometry. The second one acts on the phase space $\sim \R^2 \times  \R^2$ of the anisotropic part of the geometry.  In sections  \ref{quantmod}, \ref{Hb9} we present  a second facet of covariant integral quantization, namely  the subsequent semi-classical analysis \`a la Klauder \cite{klauder2012} stemming naturally from its formalism, especially when the quantization is provided by overcomplete families of coherent states (CS). They are affine CS, or wavelets,  for the isotropy phase space, or tensor products of the standard  CS (i.e. Schr\"{o}dinger-Glauber), or even  density operators, as those  used in the present work,  for the anisotropy phase space. This semi-classical analysis or quantum phase-space portrait is implemented in the study of the quantum mixmaster dynamics within various approximations in Section  \ref{Hb9}.  Some numerical solutions are given in Section \ref{sec:Qdynamics} and their physical features are discussed. We conclude this review by discussing possible future developments in Section \ref{perspec}. 
\section{The classical model}
 \label{classmod}
 
 \subsection{General features of the Mixmaster model}
In this section we briefly recall the Hamiltonian formulation of the Bianchi Type IX model (see e.g. \cite{Uggla}). We start from the line element on $\mathcal{M}\simeq\mathbb{S}^3\times\mathbb{R}$:
\begin{equation}
\ud s^2= -\mathcal{N}^2\ud\tau^2+\sum_ia_i^2(\omega^i)^2\, ,
\end{equation}
where the spatial one-forms $\omega_i$'s satisfy the Maurer–Cartan equation, {$\ud \omega_i=\frac{1}{2} \mfn\varepsilon_{i}^{\, jk}\omega_j \wedge \omega_k$. The lapse $\mathcal{N}(\tau)$ and the scale factors $a_i(\tau)$ are functions of time only. The respective Hamiltonian constraint in the Misner variables reads:
\begin{equation}\label{con}
\mathrm{C}=\frac{\mathcal{N}e^{-3\Omega}}{24}\left(\frac{\mathcal{V}_0}{2\kappa}\right)\left(\left(\frac{2\kappa c}{\mathcal{V}_0}\right)^2[-p_{\Omega}^2+\mathbf{ p}^2]+36\frak{n}^2e^{4\Omega}[V(\bsb)-1]\right)\,,~~(\Omega,p_{\Omega},\bsb, \mathbf{p})\in\mathbb{R}^6,
\end{equation}
where $\bsb:=(\beta_{+},\beta_-)$ and $\mathbf{p}:= (p_+,p_-)$ are canonically conjugate variables, $\mathcal{V}_0=\frac{16\pi^2}{\frak{n}^3}$ is the fiducial volume, $\kappa=8\pi Gc^{-4}$ is the gravitational constant, $\mathcal{N}$ is a non-vanishing and otherwise arbitrary function, and $\frak{n}$ is a free constant. From now on we set  $c=1$, $\frak{n}=1$ and $2\kappa=\mathcal{V}_0$. It is worthwhile noticing that the gravitational Hamiltonian $\mathrm{C}$ resembles the Hamiltonian of a particle in the 3D Minkowski spacetime and moving in a time-dependent potential. The ``Minkowskian coordinates" used in Eq. \eqref{con} have the cosmological interpretation given by the relations: 
\begin{equation}\Omega=\frac13\ln a_1a_2a_3,~~\beta_+=\frac16\ln\frac{a_1a_2}{a_3^2},~~\beta_-=\frac{1}{2\sqrt{3}}\ln\frac{a_1}{a_2}~.\end{equation}
Thus, the variable $\Omega$ describes the isotropic geometry, whereas $\beta_{\pm}$ describe distortions to the isotropic geometry and are referred to as the anisotropic variables. The potential that drives the motion of the particle represents the spatial curvature $^3R$.

Following our previous papers we canonically transform the isotropic variables,
\begin{align}
q=e^{\frac{3}{2}\Omega},~~p=\frac{2}{3}e^{-\frac{3}{2}\Omega}p_{\Omega},
\end{align}
and we choose $\mathcal{N}=-24$.
The Hamiltonian constraint (\ref{con}) is a sum of the isotropic and anisotropic parts, $\mathrm{C}=\mathrm{C}^{iso}-\mathrm{C}^{ani}_q$, where 
\begin{align}
 \label{condec1}\mathrm{C}^{iso}&=\frac{9}{4}p^2+36q^{\frac{2}{3}},\\
\label{condec2}\mathrm{C}^{ani}_q&=\frac{\mathbf{p}^2}{q^2}+36q^{\frac{2}{3}}V(\bsb)\, ,
\end{align}
and
\begin{equation}\label{b9pot}
V(\bsb) = \frac{e^{4\beta_+}}{3} \left[\left(2\cosh(2\sqrt{3}\beta_-)-e^{- 6\beta_+}
\right)^2-4\right] +  1 \,.
\end{equation}
 
Notice that $q>0$ and thus, the range of the isotropic canonical variables is the half-plane.
 
 \subsection{Approaches to the anisotropy potential}
 \subsubsection{Well-known approximations}
The anisotropy potential $V(\bsb)$ of Eq. (\ref{b9pot}) is generally regarded as too sophisticated to be used in its exact form for solving the dynamics. The best known approximations to the anisotropy potential are the harmonic and steep-wall approximation, $V_h(\bsb)$ and $V_{\triangle}(\bsb)$:
\begin{align}
V_h(\bsb)&= 8 (\beta_+^2+\beta_-^2),\\
V_{\triangle}(\bsb)&= \lim_{q\to 0} \frac{1}{3} \, \exp \left[ 4 |\ln q| (\beta_++\sqrt{3} |\beta_-|) \right].
\end{align}
While providing explicitly integrable models, the above approximations have a limited range of validity, that is, the lowest and highest excited energetic states, respectively. However, for our purposes we need to be able to model the dynamics of intermediate excitations.

\subsubsection{Perturbed Toda system}

In \cite{ewa2018} we presented a new approach to the anisotropic Hamiltonian (\ref{condec2}) for a fixed value of the isotropic variable $q$, specifically we studied  
\begin{align}\label{focus}
\mathrm{C}^{ani}=\mathbf{p}^2+V(\bsb).
\end{align}
We developed a new approximation to the anisotropic Hamiltonian with the integrable 3-particle Toda system. It smoothly approximates the three exponential walls of $V(\bsb)$ while removing the three canyons that seem responsible for the classical chaotic behavior. Furthermore, we showed that  {our quantization procedure} may quite naturally smooth out the potential in such a way as to bring to the fore the underlying Toda system and suppress the non-integrable canyons. 

We decompose the anisotropy potential as follows
\begin{align}\nonumber
V(\bsb)&=\frac{1}{3}\left(e^{4\sqrt{3}\beta_-+4\beta_+}+e^{-4\sqrt{3}\beta_-+4\beta_+}+e^{-8\beta_+}\right)\\\label{split}&-\frac{2}{3}\left(e^{-2\sqrt{3}\beta_--2\beta_+}+e^{2\sqrt{3}\beta_--2\beta_+}+e^{4\beta_+}\right)+1\\\nonumber &=V_T+V_{p}+1.
\end{align}
The introduction of new variables $q_1,~q_2,~q_3$ such that $q_1-q_2=4\sqrt{3}\beta_-+4\beta_+$ and $q_2-q_3=-4\sqrt{3}\beta_-+4\beta_+$ leads to
\begin{align}\label{newq}
V_T=\frac{1}{3}\left(e^{q_1-q_2}+e^{q_2-q_3}+e^{q_3-q_1}\right),~~
V_p=-\frac{2}{3}\left(e^{-\frac{1}{2}(q_1-q_2)}+e^{-\frac{1}{2}(q_2-q_3)}+e^{-\frac{1}{2}(q_3-q_1)}\right).
\end{align}
Lifting the above coordinate transformation to the phase space (see appendix \ref{AppT} for details) and complementing it with a rescaling of variables, $q_i\rightarrow \lambda q_i$, $p_i\rightarrow \lambda^{-1} p_i$ and $t\rightarrow 3e^{-\lambda} t$ such that $3e^{-\lambda}=\lambda^{2}$, brings the Hamiltonian (\ref{focus}) to the following form (up to an irrelevant constant):
\begin{align}\label{focus3}
\mathrm{H}=\frac{1}{2}(p_1^2+p_2^2+p_3^2)+e^{q_1-q_2}+e^{q_2-q_3}+e^{q_3-q_1}+3e^{-\frac{1}{2}\lambda}V_p.
\end{align}
The above Hamiltonian describes the periodic 3-particle Toda system \cite{berry76} plus another 3-particle Toda potential $3e^{-\frac{1}{2}\lambda}V_p$. 

 {The periodic Toda system is a system of $N$ equal-mass particles interacting via exponential forces, described by the Hamiltonian:
\begin{equation}
H=\frac12\sum_{k=1}^N p_k^2 +\sum_{k=1}^N e^{-(q_k-q_{k+1})}
\end{equation}
with periodicity condition $q_0 \equiv q_N$ and $q_1\equiv q_{N+1}$.} The periodic 3-particle Toda system is the simplest nontrivial crystal consisting of three particles, see Fig. \ref{3Toda}. It is known that the Toda systems are integrable {\cite{Henon,Flaschka,Ford}} and solutions can be derived. This system has three independent conserved quantities: the total momentum $P=p_1+p_2+p_3$, the total energy $H$ and an additional third invariant:
\begin{equation}
 K=-p_1 p_2 p_3+ p_1e^{-(q_3-q_2)}+p_2 e^{-(q_1-q_3)}+p_3e^{-(q_2-q_1)}\ .
 \end{equation}
 
In  \cite{ewa2018} we developed this approximation at the quantum level and the obtained results are summarized in section \ref{secWHIQMix}.

\section{Quantization and semi-classical formula: general features}
 \label{quantmod}
 In this section we describe our approach to quantization of the mixmaster universe based the methods of integral covariant quantizations which generalize and extend the range of applicability of the well-known `canonical quantization'.
\subsection{ What is quantization?}
 When quantizing a classical model one needs to remember that quantization is necessarily an ambiguous procedure constrained by few requirements. The ultimate validation has to be always provided by experiment. On the other hand, the latter is never sufficient to fix a unique quantization. Therefore, the concept of the unique or, ideal, quantization does not exist. The usual requirements are implied by the postulates of quantum mechanics and include linearity, the existence of a classical limit, etc. In the domain of singular gravitational systems, one often adds another requirement, namely that classically singular motions should be replaced with unitary and nonsingular ones. This, however, does not encompass the whole of what could be assumed about quantization. The important guiding principle states that quantization should respect the kinematical symmetry of the quantized model. The name given to such quantizations is covariant integral quantization and it includes coherent state quantization as a special case. As explained below, when applied to the mixmaster model it naturally yields a repulsive potential in the Hamiltonian which prevents the isotropic geometry from reaching the singularity and naturally emphasizes the underlying role of the Toda system for the anisotropic oscillations.
  
\subsection{Integral quantization}
\label{sec:intq}
Integral quantization \cite{hb2014, alijp2013,jpM2016,jpH2015,hb2017,gazeaufop18,jphb2018,gakono19} is a generic name of approaches to quantization based on operator-valued measures. It includes the so-called Berezin-Klauder-Toeplitz quantization, and more specifically  coherent state quantization \cite{alijp2013, perelomov1986}. The integral quantization framework includes as well quantizations based on Lie groups. In the sequel we will refer to this case as \textit{covariant integral quantization}. The most famous example is the covariant integral quantization based on the Weyl-Heisenberg group (WH), like the most familiar  Weyl-Wigner \cite{weyl1928, grossman1976, daub1980a, daub1980b, daub1983} and (standard) coherent states quantizations \cite{perelomov1986}. It is well established that the WH group underlies the canonical commutation rule (CCR), a paradigm of quantum physics. Actually, there is a world of quantizations that satisfy this CCR \cite{hb2014,hb2017,gazeaufop18,jphb2018}. 
Covariant integral quantizations include a more unusual quantization of the half-plane based on the affine group \cite{hb2014, jpM2016, jphb2018}. The latter is essential for our approach to quantum cosmology \cite{QC2014, QC2015, berczgamapie15A, berczgama16B} described below. A pedagogical presentation of the procedure is found in \cite{albegasca18}. See also \cite{frionalm19} for an interesting application to the quantum Brans-Dicke model.  Let us notice that  the affine group and related coherent states were also used for quantization of the half-plane in the previous works by  Klauder, although by a different method (see \cite{klauder1970, klauder2011,fanuel2013} with references therein).

The minimal requirements for a quantization are defined as follows. Given a set $X$ and a vector space $\mathcal{C}(X)$ of complex-valued functions $f(x)$ on $X$, a quantization is a linear map  $\mathfrak{Q}: f\in  \mathcal{C}(X) \mapsto \mathfrak{Q}(f) \equiv A_f \in \mathcal{A}(\mathcal{H})$ from $\mathcal{C}(X)$ to a vector space $\mathcal{A}(\mathcal{H})$ of linear operators on some Hilbert space $\mathcal{H}$. Furthermore this map must fulfill the following conditions:\\
(i) To $f=1$ there corresponds $A_f = I_\mathcal{H}$, where $ I_\mathcal{H}$ is the identity in $\mathcal{H}$,\\
(ii) To a real function $f \in \mathcal{C}(X)$ there corresponds a(n) (essentially) self-adjoint operator $A_f$ in $\mathcal{H}$.\\
Physics puts into the game further requirements, depending on various mathematical structures allocated to $X$ and $\mathcal{C}(X)$, such as a measure, a topology, a manifold, a closure etc., together with an interpretation in terms of measurements. \\
Let us assume in the sequel that $X=G$ is a Lie group with left Haar measure ${\rm d}\mu(g)$, and let $g \mapsto U_g$ be a unitary irreducible representation (UIR) of $G$ in a Hilbert space $\mathcal{H}$. Let $\mathrm{M}$ be a bounded self-adjoint operator on $\mathcal{H}$ and let us define $g$-translations of $M$ as
\begin{equation}
\label{eqMg}
\mathrm{M}(g)= U_g \mathrm{M} U_g^\dagger\,.
\end{equation}
Suppose that the  operator
\begin{equation}
\label{intgrR}
R:= \int_G  \, \mathrm{M}(g)\,\ud\mu(g) \, ,
\end{equation}
is defined in a weak sense. From the left invariance of $\ud\mu(g)$  the operator $R$ commutes with all operators $U(g)$, $g\in G$, and so, from Schur's Lemma, we have the ``resolution'' of the unity up to a constant,
\begin{equation}
\label{resunitG}
R= c_{ \mathrm{M}}I_{\mathcal{H}} \,.
\end{equation}
The constant $c_{ \mathrm{M}}$ can be found from the formula
\begin{equation}
\label{calcrho}
c_{ \mathrm{M}} = \int_G  \, \mathrm{tr}\left(\rho_0\, \mathrm{M}(g)\right)\, \ud\mu(g)\, ,
\end{equation}
where $\rho_0$ is a given unit trace positive operator. $\rho_0$ is  chosen, if manageable,  in order to make the integral convergent.  Of course, it is possible that no such finite constant exists for a given $\mathrm{M}$, or worse, it can not exist for any $\mathrm{M}$ (which is not the case for square integrable representations).
Now, if $c_{ \mathrm{M}}$ is finite and positive, the true resolution of the identity follows:
\begin{equation}
\label{Resunityrho}
\int_G \, \mathrm{M}(g) \,\ud \nu(g) = I_{\mathcal{H}}\,, \quad \ud \nu(g):= \ud\mu(g)/c_{ \mathrm{M}}\, .
\end{equation}
For instance, in the case of a square-integrable unitary irreducible representation $U: g \mapsto U_g$, let us pick a unit vector $| \psi \rangle$ for which $c_{\mathrm{M}} = \int_G {\rm d}\mu(g) |\langle \psi | U_g \psi \rangle |^2 < \infty$, i.e $| \psi \rangle$ is an admissible unit vector for $U$. With $\mathrm{M} = |\psi \rangle \langle \psi |$ the resolution of the identity (\ref{Resunityrho}) provided by the family of states 
$| \psi_g \rangle = U_g | \psi \rangle$ reads
\begin{equation}
 \int_G |\psi_g \rangle \langle \psi_g | \frac{{\rm d} \mu(g)}{c_{\mathrm{M}}} = I_{\mathcal{H}}  \,.
 \end{equation}
Vectors $| \psi_g \rangle$ are named (generalized) coherent states (or wavelet) for the group $G$. \\
 The equation (\ref{Resunityrho}) provides an integral quantization of complex-valued functions on the group $G$ as follows
 \begin{equation}
 \label{quantiz}
 f \mapsto A_f = \int_G \mathrm{M}(g) f(g) \frac{{\rm d} \mu(g)}{c_{\mathrm{M}}} \,.
 \end{equation}
 Furthermore, this quantization is covariant in the sense that:
 \begin{equation} 
 U_g A_f U_g^\dagger = A_F \quad \text{where} \quad F(g') = (\mathcal{U}_g f)(g') = f(g^{-1} g')\,,
 \end{equation}
  i.e. $\mathcal{U}_g: f \mapsto F$ is the regular representation if $f \in L^2(G, {\rm d}\mu(g))$.\\

\subsection{Integral quantization and semi-classical formula: phase-space portraits}
 \label{semclassmod}
 Integral quantization allows to develop a natural semi-classical framework. If $\rho$ and $\tilde{\rho}$ are two positive unit trace operators and furthermore if the operator $M$ of section \ref{sec:intq} above verifies $M=\rho$, we obtain the exact classical-like expectation value formula
\begin{equation}
\label{semiclass1}
{\rm tr}(\tilde{\rho} A_f) = \int_G f(g) w(g) \frac{{\rm d} \mu(g)}{c_M} 
\end{equation}
where, up to the coefficient $c_M$, $w(g) = {\rm tr}(\tilde{\rho} M(g)) \ge 0$ is a classical probability distribution on the group. Furthermore we obtain a generalization of the Berezin or heat kernel transform on $G$:
\begin{equation}
\label{semiclass2}
f \mapsto \widecheck{f}(g) = \int_G {\rm tr}(\tilde{\rho}_g \rho_{g'} ) f(g') \frac{{\rm d} \mu(g)}{c_M} 
\end{equation}
where $\tilde{\rho}_g \equiv M(g)$ when $M = \tilde{\rho}$ and $\rho_{g'} \equiv M(g')$ when $M = \rho$. The map $f \mapsto \widecheck{f}$ is a generalization of the Segal-Bargmann transform \cite{stenzel1994}. Furthermore, the function or lower symbol $\widecheck{f}$ may be viewed as a semi-classical portrait of the operator $A_f$. In the case of coherent states $|\psi_g \rangle$ (i.e. $M=\rho = |\psi \rangle \langle \psi |$), Eq.(\ref{semiclass1}) reads
\begin{equation}
\label{semiclass3}
{\rm tr}(\tilde{\rho} A_f) = \int_G f(g) \, \langle \psi_g | \tilde{\rho} | \psi_g \rangle \frac{{\rm d}\mu(g)}{c_M}\,,
\end{equation}
where $w(g) := \langle \psi_g | \tilde{\rho} | \psi_g \rangle \ge 0$ acts as a classical probability distribution on the group (up to the coefficient $c_M$). Similarly assuming $\tilde{\rho} =  |\tilde{\psi} \rangle \langle \tilde{\psi} |$, the lower symbol $\widecheck{f}(g)$ involved in (\ref{semiclass2}) reads
\begin{equation}
\label{semiclass4}
\widecheck{f}(g) = \int_G |\langle \tilde{\psi}_g | \psi_{g'} \rangle |^2 f(g') \frac{{\rm d}\mu(g')}{c_M}
\end{equation}
This point will be developed at length for the case of the affine group.

\subsection{Affine covariant integral quantization}
\label{sec:affinegroup}
\subsubsection{General settings}
The half-plane is defined as $\Pi_+ = \{(q,p) \,|\, q>0, p \in \mathbb{R} \}$. Equipped with the multiplication law
\begin{equation}
(q,p)\cdot(q',p') = \left(q q', p+ \frac{p'}{q} \right) \,,
\end{equation}
$\Pi_+$ is viewed as the affine group Aff$_+(\mathbb{R})$ of the real line, i.e., the group of transformations $\R\ni x \mapsto ax+b$, $a>0$, $b\in \R$. The left invariant measure is  ${\rm d} \mu(q,p) = {\rm d}q {\rm d}p$. The group possesses two nonequivalent square integrable UIRs. Equivalent realizations of one of them, say, $U$, are carried on  Hilbert spaces $L^2(\mathbb{R}_+, {\rm d}x/x^\alpha)$. Nonetheless these multiple possibilities do not introduce noticeable differences. Therefore we choose in the sequel $\alpha=0$, and denote $\mathcal{H} = L^2(\mathbb{R}_+, {\rm d}x)$. The UIR of  Aff$_+(\mathbb{R})$ expressed in terms of the physical phase-space variables (q,p), acts on $\mathcal{H}$ as
\begin{equation}
U_{q,p} \psi(x) = \frac{1}{\sqrt{q}} e^{\ii p x} \psi(x/q) \,.
\end{equation}
Given a unit vector $\psi \in \mathcal{H}$, we define the Affine Coherent States (ACS) as follows
\begin{equation}
\label{acsdef}
|\psi_{q,p} \rangle = U_{q,p} | \psi \rangle\,,
\end{equation}
where $\psi$ is called the fiducial vector. 

Given such a $\psi$, let us define the following integrals 
 \begin{equation}
 \label{ccoef}
 c_\alpha = \int_0^\infty \frac{{\rm d}x}{x^{2+\alpha}} \vert\psi(x)\vert^2\,.
 \end{equation}
Using the framework of covariant integral quantization presented above, we first notice that the following resolution of the identity holds
\begin{equation}
\int_{\Pi_+}  |\psi_{q,p} \rangle \langle \psi_{q,p} |\,  \frac{{\rm d}q {\rm d}p}{2 \pi c_{-1}} = I_{\mathcal{H}} \,,
\end{equation}
provided that $c_{-1} = \int_0^\infty |\psi(x)|^2 {\rm d}x /x < \infty$. Therefore the covariant integral quantization based on coherent states (ACS quantization) follows:
\begin{equation}
\label{affineq}
f \mapsto A_f = \int_{\Pi_+} f(q,p) \, |\psi_{q,p} \rangle \langle \psi_{q,p} | \,\, \frac{{\rm d}q {\rm d}p}{2 \pi c_{-1}}
\end{equation}
Note that the idea of using in quantum gravity an affine quantization instead of the Weyl-Heisenberg one was already present in Klauder's work \cite{klauder1970} devoted to the question of singularities in quantum gravity (see \cite{klauder2011} for recent references). 

\subsubsection{Properties of ACS quantization}
 In the sequel let us assume without loss that the fiducial function $\psi$ is a \underline{real} function of rapid decrease on $\mathbb{R}_+$. This ensures the convergence of the different integrals $c_\alpha$ defined in \eqref{ccoef}.
Note that  the normalization of $\psi$ corresponds to $c_{-2} = 1$. 

The first interesting issue of the map (\ref{affineq}) is that the quantization yields canonical commutation
rule, up to a scaling factor, for $A_q$ and $A_p$:
\begin{equation}
A_p = P = - \ii \frac{{\rm d}}{{\rm d}x}, \quad A_q = (c_0/c_{-1}) Q, \quad Q \psi(x)
=  x \psi(x), \quad [A_q, A_p] = \ii \frac{c_0}{c_{-1}} I_{\mathcal{H}}
\end{equation}
Through the unitary rescaling of the fiducial vector $\psi(x) \mapsto \lambda^{-1/2} \psi(x/\lambda)$ with $\lambda = c_0/c_{-1}$ we can impose $c_0 =c_{-1}$ and then recover the CCR. To simplify expressions we assume this condition to be fulfilled in the sequel.\\
However, while $A_q=Q$ is (essentially) self-adjoint, we know from \cite{reed} that $A_p=P$ is symmetric but has no self-adjoint extension. The quantization of any power of $q$ is canonical, up to a scaling factor:
\begin{equation}
A_{q^\beta} = \frac{c_{\beta-1}}{c_{-1}} Q^\beta\,.
\end{equation}
Note that our assumption on the rapid decrease of $\psi$ ensures the finiteness of the coefficients $c_{\beta-1}$, whatever $\beta$.\\
The quantization of the product $qp$ yields
\begin{equation}
A_{qp} = \frac{1}{2} (Q P + P Q) \equiv D\,,
\end{equation}
where $D$ is the dilation generator. As one of the two generators (with $Q$) of the UIR $U$ of the affine group, it is essentially self-adjoint.\\
The last and the main result  is a regularization of the quantum ``kinetic energy'':
\begin{equation}
\label{kinetic}
A_{p^2} = P^2+ \mathrm{k}_{\psi} \, Q^{-2} \quad \textrm{with} \quad \mathrm{k}_{\psi} = \int_0^\infty \frac{{\rm d}u}{c_{-1}} \, u \, (\psi'(u))^2.
\end{equation}
Therefore this quantization procedure yields a non-canonical additional term. This term is a repulsive, centrifugal like, potential whose strength depends on the fiducial vector only. In other words, this affine quantization forbids a quantum free particle moving on the positive line to reach the origin. Now, it is known \cite{reed, gesztesy1985} that the operator $P^2 = - {\rm d}^2 / {\rm d}x^2$ alone in $L^2(\mathbb{R}^+, {\rm d}x)$ is not essentially self-adjoint whereas the regularized operator (\ref{kinetic}) is for $\mathrm{k}_{\psi} \ge 3/4$. It follows that for $\mathrm{k}_{\psi} \ge 3/4$ the quantum dynamics is unitary during the entire evolution, in particular in the passage from the motion towards $x = 0$ to the motion away from $x = 0$. 

\subsubsection{Affine semi-classical portrait}
\label{sec:sIntemiclass}
The semi-classical framework sketched above applies naturally for the half-plane viewed as the affine group. The quantum states and their dynamics have phase space representations through wavelet symbols. 
To apply the map (\ref{semiclass4}) yielding lower symbols from classical $f$ we introduce two different real fiducial functions $\psi$ and $\tilde{\psi}$. $\psi$ is used for quantization and $\tilde{\psi}$ for semi-classical formula. \footnote{For technical reasons (see for instance \cite{berczgamapie15A,berczgama16B}) the vector $\psi$ is submitted to the constraints $c_{-2} = 1$, $c_0 = c_{-1}$, while $\tilde{\psi}$ is only constrained by the normalization $\tilde{c}_{-2}=1$. The coefficients $c_\alpha$ are defined in \eqref{ccoef}.} 
The map (\ref{semiclass4}) yields in the present case:
\begin{equation}
f(q,p) = q^\beta \mapsto \widecheck{f}(q,p) = \frac{\tilde{c}_{-\beta-2} \, c_{\beta-1}}{c_{-1}} q^\beta \, ,
\end{equation}
where $c$ coefficients (defined in \eqref{ccoef}) stand for $\psi$ and $\tilde{c}$ coefficients for $\tilde{\psi}$. \\
We notice that $\widecheck{q} = c_0\, \tilde{c}_{-3} \, (c_{-1})^{-1} q = \tilde{c}_{-3} \, q$. Therefore we must impose $\tilde{c}_{-3} = 1$ in order to obtain for physical consistency $\widecheck{q} = q$. This constraint is obtained through a simple rescaling of the fiducial vector $\tilde{\psi}$. If  this condition is fulfilled, other important symbols are
\begin{align}
f(q,p) = p & \mapsto \widecheck{f}(q,p) = p\\ \label{LSofHAM}
f(q,p) =p^2 &\mapsto \widecheck{f}(q,p) = p^2+ \mathrm{k}_s(\tilde{\psi},\psi) \, q^{-2}\\
f(q,p) = q p & \mapsto \widecheck{f}(q,p) = q p
\end{align}
where
\begin{equation}\label{SofKs}
\mathrm{k}_s(\tilde{\psi},\psi) = \int_0^\infty (\tilde{\psi}'(u))^2 \, {\rm d}u + \tilde{c}_0\, \mathrm{k}_\psi \,.
\end{equation}

\subsection{Weyl-Heisenberg integral quantization}
\label{secWH}
\subsubsection{General settings}
The plane is denoted by $\Pi = \{ (q ,p) \,|\, q, p \in \mathbb{R} \} \equiv \mathbb{R}^2$. Equipped with the law of addition, $\Pi $ is viewed as the abelian translation group $\mathbb{R}^2$. 
 This leads naturally to the unique (up to equivalence) unitary irreducible projective representation $(q,p)\mapsto U(q,p)$ of the Weyl-Heisenberg group on $\mathcal{H} = L^2(\mathbb{R}, {\rm d}x)$:
 \begin{equation}
\label{WHU}
\begin{split}
U(0,0) &= I\, , \quad  U^{\dag}(q,p) = U(-q,-p)\, , \\
 U(q,p)\,U(q^{\prime},p^{\prime})& = e^{\mathrm{i} \xi\left((q,p),(q^{\prime},p^{\prime})\right)}U(q + q^{\prime}q, p + p^{\prime})\, , 
\end{split}
\end{equation}
where the real valued $\xi$ encodes the non commutativity of the representation, i.e.,  the central feature of the quantization. It has   to fulfill  cocycle conditions which correspond with group structure of $\mathbb{R}^2$. Therefore the unique choice, besides the trivial one, reads  $U(q,p) =  e^{\mathrm{i} (p\hat q-q\hat p)}$. $U(q,p)$} is the unitary displacement operator where $[\hat{q}, \hat{p}] = \mathrm{i} I_{\mathcal{H}}$ and $\xi\left((q,p),(q^{\prime},p^{\prime})\right)$ is the symplectic form:
\begin{equation}
\xi\left((q,p),(q^{\prime},p^{\prime})\right) = k\,(qp^{\prime}-q^{\prime}p)\, . 
\end{equation}
Here  $k$ is a parameter that quantum physics fixes to $1/\hbar$, and  for convenience it is  put equal to $1$ in these considerations. Moreover, from the translational invariance the invariant measure is $\mathrm{d} q\,\mathrm{d} p$. The operator $M\left(q,p\right) \equiv M_g$ of \eqref{eqMg} reads
\begin{equation}
\label{solQ}
M\left(q,p\right) := U(q,p)\,M \,U(q,p)^{\dag} \, .
\end{equation}
The choice of  $M$ is admissible provided that  the parameter $c_M$ introduced in section \ref{sec:intq} is finite, and if $M$ is trace class.
The family of operators $M\left(q,p\right)$ solves the identity  in $\mathcal{H}$ according to:
\begin{equation}
\label{1A1}
\int_{\mathbb{R}^2} M(q,p)\, \frac{\mathrm{d}q\,\ud p}{c_{M}} = I_{\mathcal{H}}\,.
\end{equation}
 This yields the Weyl-Heisenberg covariant integral quantization which transforms a function (more generally a distribution) $f(q,p)$ into an operator $A_f$ acting on $\mathcal{H}$ through the linear map
\begin{equation} \label{fAf}
f(q,p) \mapsto A_f = \int_{\mathbb{R}^2} f(q,p)\, M(q,p)\, \frac{\ud q\,\ud p}{c_{M}}\, . 
\end{equation}
Translational covariance  holds in the sense that the quantization of the translation of $f$ is unitarily equivalent to the quantization of $f$  as follows:
\begin{equation} \label{covtrans1}
U(q_0,p_0)\,A_f \,U(q_0,p_0)^{\dag}= A_{\mathcal{T}(q_0,p_0)f}\, , \quad \left(\mathcal{T}(q_0,p_0)f\right)(q,p):= f\left(q-q_0,p-p_0\right) \, . 
\end{equation} 
Let us end this section by presenting an alternative form  of the Weyl-Heisenberg quantization \eqref{fAf} thanks to the ``WH-transform'' of the operator $M$ and its inverse.  Let us introduce the function on the plane:
\begin{equation}
\label{WHT}
\Pi(q,p) = \mathrm{Tr}\left(U(-q,-p)M \right)\  \Leftrightarrow\ M = \int_{\mathbb{R}^2} U(q,p) \, \Pi(q,p)\,\frac{\ud q\,\ud p}{2\pi}\,. 
\end{equation}
The inverse WH-transform exists due to remarkable properties of the displacement operator  $U(q,p)$ \cite{gazeaufop18}. Therefore we have at our disposal  the equivalent formulation of the Weyl-Heisenberg integral quantization based on the so-called symplectic Fourier transform:
 \begin{equation}
\label{symFourqp}
 \mathfrak{F_s}[f](q,p)= \int_{\mathbb{R}^2}e^{-\mathrm{i}\left(qp^{\prime}-pq^{\prime}\right)}\, f(q^{\prime},p^{\prime})\,\frac{\ud q^{\prime}\,\ud p^{\prime}}{2\pi} \, . 
\end{equation}
Because $\mathfrak{F_s}$ is involutive,  the WH integral quantization of \eqref{fAf} reads:
 \begin{equation} 
 \label{SIQ}
A_f= \int_{\mathbb{R}^2}  U(q,p)\,  \overline{\mathfrak{F_s}}[f](q,p)\, \frac{\Pi(q,p)}{\Pi(0,0)} \,\frac{\ud q\,\ud p}{2\pi}\,,
\end{equation}
where $ \overline{\mathfrak{F_s}}[f](q,p):=  \mathfrak{F_s}[f](-q,-p)$. The value of the constant $c_{M}$ introduced in  \ref{sec:intq}  is  $c_{M} =2 \pi \, \mathrm{Tr}\left(M \right) = 2 \pi \Pi(0,0)$. From now on and with no loss of generality, we impose $\Pi(0,0)=1$. 
\\
Furthermore, the semi-classical portrait or lower symbol of $A_f$ is given by the integral
\begin{equation}
\label{WHlowsym}
\widecheck{f}(q,p)= \int_{\mathbb{R}^2}  \,  \mathfrak{F_s}\left[\Pi\,\widetilde{\Pi}\right](q^{\prime}-q,p^{\prime}-p)\, f(q^{\prime},p^{\prime}) \,\frac{\ud q^{\prime}\,\ud p^{\prime}}{2\pi}\, , \quad \widetilde{\Pi}(q,p):=\Pi(-q,-p)\,.
\end{equation}
\subsubsection{Properties of the Weyl-Heisenberg integral quantization}
There are several features  independent of the choice of the quantization operator $M$.
First, the canonical commutation rule is preserved
\begin{equation}
\label{AqAp}
A_q = \hat q + c_0I_{\mathcal{H}}\, , \quad A_p= \hat p+d_0I_{\mathcal{H}}\,, \quad c_0,d_0\in \mathbb{R}\, ,  \rightarrow \left[A_q,A_p\right]= \mathrm{i} I_{\mathcal{H}}\, .
\end{equation}
For the kinetic energy we have the following formula
\begin{equation}
 A_{p^2}= \hat p^2 + e_1\,\hat p + e_0I_{\mathcal{H}}\, , \quad e_0, e_1 \in \mathbb{R}\, . 
\end{equation}
The constants $c_0$, $d_0$, $e_0$, $e_1$ appearing in the above can be eliminated through a suitable choice of $M$.
The quantization of the dilation  operator yields:
\begin{equation}
A_{qp} = A_q\,A_p + \mathrm{i} f_0I_{\mathcal{H}}\, , \quad  f_0\in \mathbb{R}\, . 
\end{equation}
This operator can be brought to the self-adjoint dilation operator $(\hat q\hat p + \hat p\hat q)/2$ again through a suitable choice of $M$.\\
The quantization of more involved functions of $q$ or $p$ combines in general  multiplication operators with convolutions and (pseudo-) differential operators  \cite{hb2017,gazeaufop18}.

 \subsubsection{Weyl-Heisenberg semi-classical portrait}
By choosing the separable gaussian weight function $\Pi$,
\begin{equation}
\label{sepgauss}
\Pi(q,p)=e^{-\frac{q^2}{2\sigma^2}}e^{-\frac{p^2}{2\tau^2}}\,,
\end{equation}
we obtain for the lower symbol \eqref{WHlowsym} of the quantum operator $A_f$ the integral,
 \begin{equation}
\label{WHlowsymSG}
\widecheck{f}(q,p)= \frac{\sigma \tau}{2}\int_{\mathbb{R}^2} e^{-\frac{\tau^2}{4}(q^{\prime}-q)^2}\,e^{-\frac{\sigma^2}{4}(p^{\prime}-p)^2}\, f(q^{\prime},p^{\prime}) \,\frac{\ud q^{\prime}\,\ud p^{\prime}}{2\pi}\, .
\end{equation}
In particular, we obtain the following relevant lower symbols for Bianchi IX model:
\begin{equation}
\label{WHIQexpsp}
e^{- a q} \mapsto \widecheck{(e^{-a q})}=e^{\frac{a^2}{\tau^2}}e^{-a q}\,, \quad p^2 \mapsto \widecheck{(p^2)}=p^2+\frac{2}{\sigma^2} \,.
\end{equation}

\section{Quantization of the mixmaster Hamiltonian}
\label{Hb9}
In this section we apply the quantization methods described previously to the mixmaster universe in its canonical formulation.
 
\subsection{General settings}
The classical Hamiltonian constraint $\mathrm{C}$ of \eqref{con} with its isotropic and anisotropic parts, $\mathrm{C}^{(\mathrm{iso})}$ and $\mathrm{C}^{(\mathrm{anis})}_q$ of \eqref{condec1} and \eqref{condec2} reads as
\begin{align}
\mathrm{C} &= \mathrm{C}^{(\mathrm{iso})} - \mathrm{C}^{(\mathrm{anis})} \\
\mathrm{C}^{(\mathrm{iso})} &= \frac{9}{4}p^2+36q^{\frac{2}{3}},\\
\mathrm{C}^{(\mathrm{anis})}_q&=\frac{\mathbf{p}^2}{q^2}+36q^{\frac{2}{3}}V(\bsb)\, .
\end{align}
The proposed quantization is a compound procedure that fully complies with the symmetries of the phase space: an affine integral quantization for the isotropic variables $(q,p)$ which is consistent  with the dilation-translation symmetry of the half-plane (affine group), and the  Weyl-Heisenberg covariant integral quantization (WHCIQ) for the anisotropic variables $(\beta_\pm, p_\pm)$ which is consistent with the translation symmetry of the plane (Weyl-Heisenberg group). However, for the anisotropic part,  we first proceed in \ref{sec:ACSandCanonical} with the standard canonical quantization before displaying in the subsequent \ref{secWHIQMix} the improvements 
afforded by the WHCIQ.
\subsection{ACS $\oplus$ Canonical quantization}
\label{sec:ACSandCanonical}
\subsubsection{The quantum framework}
In our previous papers \cite{berczgamapie15A, berczgamapie15B,berczgama16B} we used the ACS quantization framework presented above for the isotropic pair $(q,p)$, and  a canonical quantization for the anisotropic pairs $(\beta_\pm, p_\pm)$. We obtain the quantized version $\hat{\mathrm{H}} \equiv A_{\mathrm{C}}$ of the classical Hamiltonian $\mathrm{C}$ acting on the Hilbert space $\mathcal{H} = \mathcal{H}^{{\rm (iso)}} \otimes  \mathcal{H}^{{\rm (anis)}}$, where $\mathcal{H}^{{\rm (iso)}} =L^2(\mathbb{R}_+, {\rm d}x)$ and $\mathcal{H}^{{\rm (anis)}} =L^2(\mathbb{R}^2, {\rm d}\beta_+ {\rm d} \beta_-)$:
\begin{equation}
\label{qb9}
\left.\begin{array}{c}
\mathrm{C} \mapsto A_{\mathrm{C}} \equiv \widehat{\mathrm{C}} =  \widehat{\mathrm{C}}^{{\rm (iso)}} - \widehat{\mathrm{C}}^{{\rm (anis)}}(Q)\\
 \widehat{\mathrm{C}}^{{\rm (iso)}} \equiv A_{\mathrm{C}^{{\rm (iso)}}} = \frac{9}{4} \left(P^2+\frac{\mathrm{k}_1}{Q^2} \right)+ 36 \mathrm{k}_3 Q^{2/3}\\
\widehat{\mathrm{C}}^{(\mathrm{anis})}(q) \equiv A_{\mathrm{C}^{(\mathrm{anis})}_q}=\mathrm{k}_2\frac{\hat{p}_+^2+\hat{p}_-^2}{q^2}+
36 \mathrm{k}_3 q^{2/3} V(\hat{\beta}_+,\hat{\beta}_-)\,,
\end{array}\right.
\end{equation}
where $\hat{p}_\pm = -\ii \partial_{\beta_\pm}$, and the positive coefficients $\mathrm{k}_1$, $\mathrm{k}_2$, $\mathrm{k}_3$ result from our ACS quantization, being only dependent on the ACS fiducial vector (see appendix \ref{appendA}). This precisely represents  the main interest of our ACS quantization pointed out in the section \ref{sec:affinegroup}, namely the appearance  of a repulsive potential $\mathrm{k}_1 Q^{-2}$ which yields,  within the Bianchi IX framework,  the resolution of the singularity.\\
Furthermore,   a thorough study of the Hamiltonian $\widehat{\mathrm{C}}^{(\mathrm{anis})}(q)$ shows  that despite three open canyons, the potential $V(\beta_+,\beta_-)$ originates a purely discrete spectrum \cite{berczugamal17}. Therefore the operator $\widehat{\mathrm{C}}^{(\mathrm{anis})}(q)$ has the discrete spectral resolution
\begin{equation}
\label{spectralres}
\widehat{\mathrm{C}}^{(\mathrm{anis})}(q)  = \sum_n E_n^{{\rm (anis)}} (q) | e_n^{{\rm (anis)}} (q) \rangle \langle e_n^{{\rm (anis)}} (q) | \,.
\end{equation}
We prove in \cite{berczugamal17} that the eigenenergies $E_n(q)$ verify
$
\lim_{q \to 0} q^2 \, E_n(q) = 0\,.
$
This property is crucial for proving the resolution of the classical singularity within the framework of  adiabatic approximations to the quantum behavior of the system.\\
Finally we introduce a unitary transformation $U(q,q')$ and a new self-adjoint operator $\hat{\mathrm{A}}(q)$ acting on the Hilbert space $\mathcal{H}^{{\rm (anis)}}$. They will be useful for studying in Section \ref{sec:Qdynamics} the quantum behavior:
\begin{equation}
\label{newunitary}
U(q,q') = \sum_n | e_n^{{\rm (anis)}} (q) \rangle \langle e_n^{{\rm (anis)}} (q') | \,.
\end{equation}
We notice that $U(q,q')^\dagger = U(q',q)$ and $U(q,q) = I_{\mathcal{H}^{{\rm (int)}}}$.
We also define  the self-adjoint operator $\hat{\mathrm{A}}(q)$   as
\begin{equation}
\label{Afield}
\hat{\mathrm{A}}(q) = \ii \sum_n \left( \frac{\partial}{\partial q}  | e_n^{{\rm (anis)}} (q) \rangle \right)  \langle e_n^{{\rm (anis)}} (q)| = \ii \left( \frac{\partial}{\partial q} U(q,q') \right) U(q',q) \,.
\end{equation}

In Section \ref{sec:adiabvibro} we analyze  the quantum dynamical properties obtained in \cite{berczgamapie15A, berczgamapie15B,berczgama16B} for this system.

\subsection{ACS $\oplus$ Covariant Weyl-Heisenberg integral quantization}
\label{secWHIQMix}
In this approach the quantization of the isotropic part is unchanged from the previous case, but we focus on the anisotropic part by replacing the canonical quantization with WHCIQ.
\subsubsection{The framework}
\begin{figure}[!ht]
\center
\begin{tabular}{cc}
\includegraphics[width=0.35\textwidth]{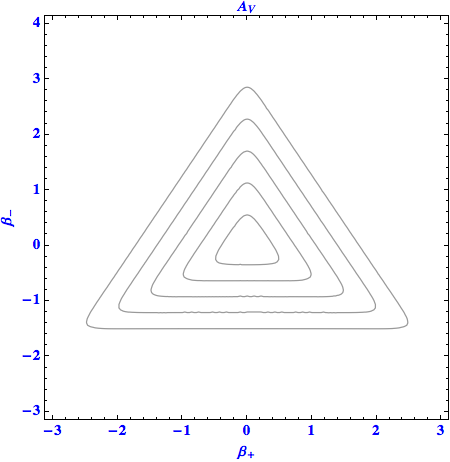}
\end{tabular}
\caption{Contour plot  of the symmetric quantized Bianchi IX potential near its minimum, $D=1.2$. }
\label{figure2}
\end{figure}

In \cite{ewa2018} we applied the integral quantization method described in section \ref{secWH} to the anisotropy potential \eqref{b9pot}. 
The selected  function $\Pi$ was of the separable Gaussian type \eqref{sepgauss} 
\begin{equation}
\Pi(\beta_+,p_+;\beta_-,p_-)=e^{-\frac{\beta_+^2}{2\sigma_+^2}}\,e^{-\frac{\beta_-^2}{2\sigma_-^2}}\,e^{-\frac{p_+^2}{2\tau_+^2}}\,e^{-\frac{p_-^2}{2\tau_-^2}}\,.
\end{equation}
Now, with the weight function $\Pi(q,p)= e^{-\frac{q^2}{2\sigma^2}}\,e^{-\frac{p^2}{2\tau^2}}$ and the original function depending on $q$ only,  $f(q,p)=u(q)$, the quantization map \eqref{SIQ} reads:
\begin{equation}
\label{OpAuq}
u(q) \mapsto {A}_{u}=\frac{\tau}{\sqrt{2\pi}}\left(u\ast e^{\frac{\tau^2(\cdot)}{2}}\right)(\hat q)\,,
\end{equation}
Since the Bianchi IX potential is written as a sum of products of exponentials, the application  of \eqref{OpAuq}   
to the exponential case $u(q)=e^{a q}$ yields:
\begin{equation}
e^{a q}\mapsto {A}_{e^{a q}}=e^{\frac{a^2}{2\tau^2}}e^{a \hat q}\,.
\end{equation}
Applied separately to each of the  configuration variables $\beta_{\pm}$, this map yields the multiplication operator
\begin{align}
V(\beta_+,\beta_{-})\mapsto {A}_{V(\beta_+,\beta_{-})}=\frac13&\left(2D^{16} e^{4\beta_+}\cosh 4\sqrt{3}\beta_--4
D^4 e^{-2\beta_+}\cosh 2\sqrt{3}\beta_-\right.
\nonumber\\ &\left.+D^{16} e^{-8\beta_+}-2D^4 e^{4\beta_+}\right) +1, \label{isopot} 
\end{align}
where  we omit hats in $\hat \beta_{\pm}$ for the sake of simplicity and $D= e^{\frac{2}{\sigma^2}}$. The classical anisotropy potential $V(\beta_+,\beta_{-})$ is recovered for $D=1$ (or, $\sigma \rightarrow \infty$).
The new potential is shown in Fig. \ref{figure2}. One may verify that it is invariant with respect to the rotations by $2\pi/3$ and $4\pi/3$, and thus, the $\mathrm{C}_{3v}$ symmetry of the initial Bianchi IX potential is preserved in the full plane.

\subsubsection{ Underlying Toda system}
The quantized potential \eqref{isopot}, similarly  to the classical potential in Eq. (\ref{split}), may be written as a difference between two Toda potentials ({recall} that $V_p$ is negative):
\begin{align}\label{isopot1}
 {A}_{V(\beta_+,\beta_{-})}=&D^{16}V_T+D^{4}V_p+1,
\end{align}
where the coefficients $D^{16}$ and $D^{4}$ strengthen the dynamical role of $V_T$ as $D>1$. 
We show in  \cite{ewa2018} that the ratio $\left|\frac{{V}_{p}}{{V}_{T}} \right|$ is such that  $\left|\frac{{V}_{p}}{{V}_{T}} \right| \le 2$. Therefore for $D \gg1$, the exact quantum potential may be viewed as a  perturbed Toda potential $D^{16}V_T$. This brings new possibilities to the study of the Mixmaster by means of the periodic 3-body Toda system. The latter has been analyzed both on the classical and quantum levels. In the literature, one may find ways to construct classical solutions \cite{Flaschka,KacMoerbeke} as well as the corresponding eigenfunctions and eigenvalues \cite{Gutzwiller,VanMoerbeke}.  Thus, we have provided integrable and analytically solvable approximation to the anisotropic potential of the Bianchi IX model not only in the IR and UV limits, as the usual harmonic and steep-wall approximations do, but also in the vast, unexplored in-between region, on both classical and quantum levels.

\section{Quantum dynamical studies}
\label{sec:Qdynamics}
 The previous section \ref{Hb9} dealt with the question of quantization of the mixmaster universe. In the present section we describe tools for the analysis of its quantum motion developed within the quantum framework of section \ref{sec:ACSandCanonical}.  They enable us to  get insight into the rich physics of the quantum mixmaster bounce.

 The quantum dynamics of mixmaster was analysed in the works \cite{berczgamapie15A,berczgamapie15B,berczgama16A,berczgama16B}, where  it was  shown how to remove the classical singularity by means of quantization and how to use approximation methods to solve the quantum dynamics. We first employed affine coherent states to establish a semi-classical description of the quantum dynamics of the isotropic degrees of freedom and then we used adiabatic approximations to derive solutions to the quantum equations of motion. The found solutions correspond to the quantum Friedmann-like universes in which the dynamics of isotropic variables is fuelled by the energy of a fixed eigenstate of the anisotropic Hamiltonian. However, these solutions are very special. We subsequently developed a nonadiabatic approximation and used it to derive more accurate equations of motion to which we found more solutions. It turned out that the quantum bounce generically involves complex interplay between isotropic and anisotropic variables, which leads to a very rich dynamics of the quantum universe.  The mixmaster bounce is typically very asymmetric in time. Notably, the bounce is immediately followed by an extended phase of accelerated expansion that can last for an arbitrarily long time. We found that the more matter and the more anisotropy there is in the universe, the smaller the volume which the universe reaches at the bounce, the more sudden or stiff the bounce is and the longer the post-bounce inflationary phase lasts. We have showed that plenty of post-bounce accelerated expansion should occur in a realistic cosmological scenario. In what follows we explain in some detail how those results were derived.

\subsection{Semi-classical Lagrangian and dynamical equations}
\label{sec:semilagrange}
Analogously to  the so-called ``enhanced quantization'' promoted  by Klauder (see for instance  \cite{klauder2012} and references therein), we developed a compound semi-classical Lagrangian approach of the exact quantum Hamilonian of section \ref{sec:ACSandCanonical}: semi-classical for the isotropic variable and purely quantum for the anisotropy variables.
We also expanded the anisotropy potential about its minimum in order to deal with its harmonic approximation suitable for both analytical and numerical treatments.
Moreover, following standard approaches in quantum molecular physics, we studied successively adiabatic (Born-Oppenheimer-like) \cite{berczgamapie15B,berczgamapie15A} and nonadiabatic (vibronic-like) \cite{berczgama16B} approximations of the quantum dynamical equations. 

We recall below our procedure detailed in \cite{berczgamapie15A,berczgama16B}. It is based on a consistent framework allowing us to approximate the quantum Hamiltonian and its associated dynamics (in the constraint surface) by making use of the semi-classical Lagrangian approach. This is made possible thanks to  our ACS formalism. The quantum constraint \eqref{qb9} has the general form
\begin{equation}
\left.\begin{array}{c}
\widehat{\mathrm{C}} =\widehat{\mathrm{C}}^{(\mathrm{iso})} - \widehat{\mathrm{C}}^{(\mathrm{anis})}(Q)\\
\widehat{\mathrm{C}}^{(\mathrm{iso})} =  \frac{9}{4}P^2+W(Q), \quad W(q) = \dfrac{9\mathrm{k}_1}{4 q^2} + 36 \mathrm{k}_3 q^{2/3}
\end{array}\right.
\end{equation}
and the $q$-dependent Hamiltonian $\widehat{\mathrm{C}}^{(\mathrm{anis})}(q)$ is formally the one of (\ref{spectralres}) that acts on the
Hilbert space of anisotropy states. The Schr\"{o}dinger equation (here $\hbar=1$)
\begin{equation}
\ii \frac{\partial}{\partial t} | \Phi(t) \rangle = N \widehat{\mathrm{C}} |\Phi(t) \rangle
\end{equation}
can be deduced from the Lagrangian
\begin{equation}
\label{qlagrangian}
\mathrm{L}(\Phi, \dot{\Phi}, N) = \langle \Phi(t) | \left( \ii  \frac{\partial}{\partial t} - {N}  \widehat{\mathrm{C}}\right)  |\Phi(t) \rangle \,,
\end{equation}
via the minimization of the respective action with respect to $ |\Phi(t) \rangle$. The quantum counterpart of the classical constraint $\mathrm{C} = 0$ can be obtained as follows:
\begin{equation}
\label{constraint}
- \frac{\partial \mathrm{L}}{\partial {N}} =  \langle \Phi(t) |  \widehat{\mathrm{C}}  |\Phi(t) \rangle =0\,.
\end{equation}
The commonly used Dirac method of imposing constraints, $ \widehat{\mathrm{C}}  |\Phi(t) \rangle =0$ implies (\ref{constraint}) but the reciprocal does not hold in general. This means that a state $ |\Phi(t) \rangle$ satisfying (\ref{constraint}) does not necessarily lie in the kernel of the operator $ \widehat{\mathrm{C}} $.\\
We assume that $ |\Phi(t) \rangle $ reads
\begin{equation}
\label{coupledstate}
\left.\begin{array}{c}
|\Phi(t) \rangle = U(Q,q_0) \left( | \tilde{\psi}_{q(t), p(t)} \rangle \otimes | \phi^{{\rm (anis)}} (t) \rangle \right) \\
| \phi^{{\rm (anis)}} (t) \rangle = \sum_n c_n(t) |e_n^{{\rm (anis)}}(q_0) \rangle \,,
\end{array}\right.
\end{equation}
where the different elements are defined as follows: \\
(a) $ | \tilde{\psi}_{q(t), p(t)} \rangle \in \mathcal{H}^{{\rm (iso)}}$ is a $(q,p)$-time-dependent ACS, the fiducial vector $\tilde{\psi}$ being constrained by $\tilde{c}_{-3}=1$ as in the section \ref{sec:sIntemiclass}, \\
(b) $U(Q,q_0)$ is the unitary operator resulting from the substitution $q \mapsto Q$ in the operator defined in (\ref{newunitary}), \\
(c) $q_0$ is an arbitrary fixed value of $q$. \\
The role of the unitary operator $U(Q,q_0)$ is to introduce a  natural entanglement (quantum coupling) between the isotropic state and the anisotropic one in such a way that asymptotically they decouple for large values of the scale factor. Therefore the coupling of states occurs  essentially during the bounce. As we will see below,  $U(Q,q_0)$, or more precisely its derivative $\hat{\mathrm{A}}(q)$ of Eq. \eqref{b9dynamical}, is responsible of nonadiabatic effects (excitations or decay of anisotropy states).\\
By replacing  $|\Phi(t) \rangle$ in (\ref{qlagrangian}) with the expression above (\ref{coupledstate}), we obtain the following semi-classical Lagrangian $\mathrm{L}^{{\rm semi}}(q,\dot{q}, p, \dot{p}, \phi^{{\rm (anis)}}, \partial_t \phi^{{\rm (anis)}}, {N})$ (see \cite{berczgama16B} for more details):
\begin{align}
\label{Newlagrangian}
\nonumber \mathrm{L}^{{\rm semi}}(q,\dot{q}, p, \dot{p},& \phi^{{\rm (anis)}}, \partial_t \phi^{{\rm (anis)}}, {N}) = -q \, \dot{p} +  \langle \phi^{{\rm (anis)}} (t) | i \frac{\partial}{\partial t} | \phi^{{\rm (anis)}} (t) \rangle \\
 - {N} \mathrm{C}_s^{{\rm (iso)}}&(q,p) +{N} \langle \phi^{{\rm (anis)}} | \widehat{\mathrm{C}}^{{\rm (anis)}}_s(q,p) | \phi^{{\rm (anis)}}  \rangle
\end{align}
To avoid introducing new unessential constants, we neglect in the sequel the dressing effects of semi-classical formula (functions of $Q$) given in  \ref{sec:sIntemiclass}. In this case the real function $\mathrm{C}_s^{{\rm (iso)}}(q,p)$ and the new operator $\widehat{\mathrm{C}}^{{\rm (anis)}}_s(q,p)$ read:
\begin{equation}
\label{semiconstr}
\left.\begin{array}{c}
\mathrm{C}_s^{{\rm (iso)}}(q,p) =  \dfrac{9}{4} p^2+ \widetilde{W}(q)\,, \quad \widetilde{W}(q) = \dfrac{\tilde{\mathrm{k}}}{q^2} + W(q) \,,\\
\widehat{\mathrm{C}}^{{\rm (anis)}}_s(q,p) = -\dfrac{9}{2}  p\, \hat{\mathrm{A}}(q) +\dfrac{9}{4}  \hat{\mathrm{A}}(q)^2 +\sum_n E_n(q)  | e_n^{{\rm (anis)}} (q_0) \rangle \langle e_n^{{\rm (anis)}} (q_0) |\,,
\end{array}\right.
\end{equation}
where $\hat{\mathrm{A}}(q)$ is the self-adjoint operator defined in (\ref{Afield}).\\
From (\ref{Newlagrangian}) and (\ref{semiconstr}) we deduce the complete set of dynamical equations including the action of the isotropic variable on the anisotropic ones and the backaction of the anisotropic variables on the isotropic one:
\begin{equation}
\label{b9dynamical}\boxed{
\left.\begin{array}{c}
\dot{q} = {N} \dfrac{\partial}{\partial p}\left( \mathrm{C}_s^{{\rm (iso)}}(q,p) - \langle \phi^{{\rm (anis)}} | \widehat{\mathrm{C}}^{{\rm (anis)}}_s(q,p) | \phi^{{\rm (anis)}}  \rangle \right)\\
\dot{p} = - {N} \dfrac{\partial}{\partial q}\left( \mathrm{C}_s^{{\rm (iso)}}(q,p) - \langle \phi^{{\rm (anis)}} | \widehat{\mathrm{C}}^{{\rm (anis)}}_s(q,p) | \phi^{{\rm (anis)}}  \rangle \right)\\
\ii \, \dfrac{\partial}{\partial t}  | \phi^{{\rm (anis)}}  \rangle = - {N} \widehat{\mathrm{C}}^{{\rm (anis)}}_s(q,p) | \phi^{{\rm (anis)}}  \rangle
\end{array}\right.}
\end{equation}
The classical constraint $\mathrm{C} = 0$ is given in this framework by the semi-classical formula
\begin{equation}
\label{NewQc}
- \frac{ \partial \mathrm{L}^{{\rm semi}}}{\partial {N}} = \mathrm{C}_s^{{\rm (iso)}}(q,p) - \langle \phi^{{\rm (anis)}} | \widehat{\mathrm{C}}^{{\rm (anis)}}_s(q,p) | \phi^{{\rm (anis)}}  \rangle = 0
\end{equation}
The Hubble rate $\textrm{H}$ from (\ref{b9dynamical}) reads
\begin{equation}
\textrm{H} = \dfrac{2}{3 {N} } \dfrac{\dot{q}}{q} = \dfrac{3}{q}  \left( p - \langle \phi^{{\rm (anis)}} |\hat{\mathrm{A}}(q)  | \phi^{{\rm (anis)}}  \rangle \right) \,.
\end{equation}
Therefore we obtain from (\ref{NewQc}) the modified Friedman equation
\begin{equation}
\left.\begin{array}{c}
\dfrac{1}{4}\textrm{H} ^2+\dfrac{9}{4 q^2} \sigma_A(q)^2 + \dfrac{\widetilde{W}(q)}{q^2}  - \sum_n \dfrac{E_n(q)}{q^2} \left| \langle  \phi^{{\rm (anis)}} | e_n^{{\rm (anis)}}(q_0) \rangle \right|^2  = 0\,,\\
\sigma_A(q)^2 = \langle \phi^{{\rm (anis)}} |\hat{\mathrm{A}}(q)^2  | \phi^{{\rm (anis)}}  \rangle-\left( \langle \phi^{{\rm (anis)}} |\hat{\mathrm{A}}(q)  | \phi^{{\rm (anis)}}  \rangle \right)^2 
\end{array}\right.
\end{equation}
where $\textrm{H} $, $q$ and $ | \phi^{{\rm (anis)}}  \rangle$ are implicitly time-dependent.\\
Since the dynamical system (\ref{b9dynamical}) does not admit explicit analytical solutions, two kinds of approximations will be investigated in the sequel.

\subsection{Adiabatic and nonadiabatic approximations}
\label{sec:adiabvibro}

In these different approximations we found that the classical singularity is always removed, being replaced with a quantum bounce. The adiabatic approximation (Born-Oppenheimer) allows to find approximate solutions that look like Friedmann models for which the dynamics of the scale factor is fueled by the eigenenergy of a \emph{fixed} eigenstate of the anisotropic Hamiltonian. Therefore this approximation prohibits by definition excitation or decay of the quantum anisotropic states. To study the excitation/decay effect we developed at a second stage the nonadiabatic (vibronic) approximation. We found that in general  strong nonadiabatic effects are involved during a quantum bounce. Notably the mixmaster bounce seems very asymmetric in time, the bounce being immediately followed by an extended phase of large excitation of anisotropy which in turn leads to an accelerated expansion \cite{berczgamapie15A}.

\subsubsection{Adiabatic (Born-Oppenheimer) approximation \cite{berczgamapie15A,berczgamapie15B}}
\begin{figure}[htb]
\center
\includegraphics[width=0.45\textwidth]{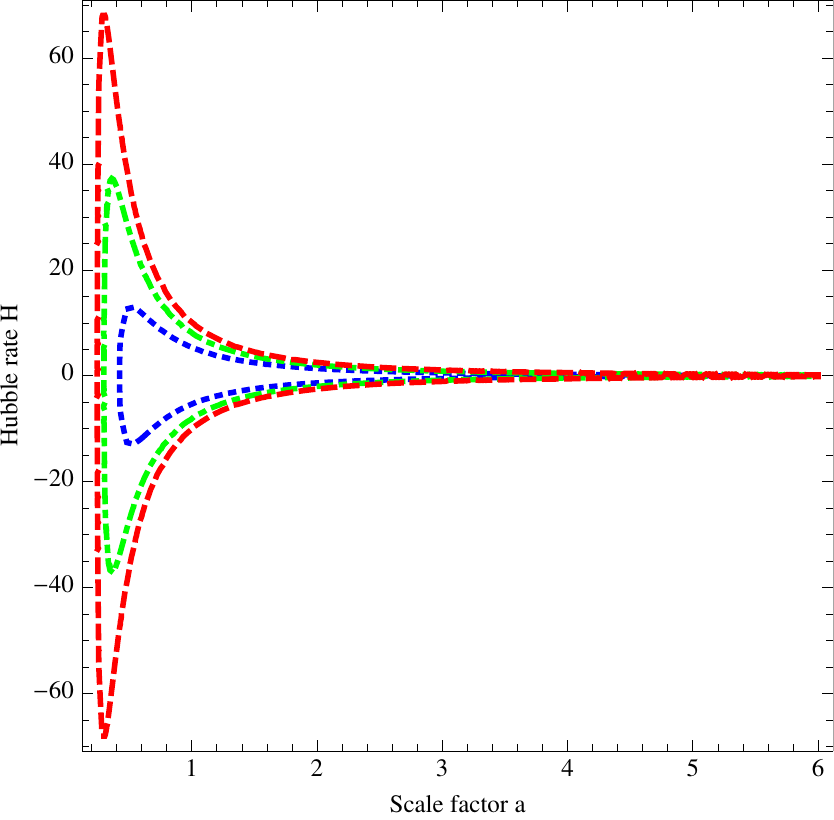} 
\caption{Adiabatic (Born-Oppenheimer-like) approximation: plot of different trajectories (i.e. different values of $n_0$) in the plane $(a = q^{2/3}, \textrm{H})$. The classical singularity is replaced by a quantum bounce. (Source: \cite{berczgamapie15A})}
\label{figBIXadiab}
\end{figure}

 A detailed analysis of (\ref{b9dynamical}) shows that only the operator $\hat{\mathrm{A}}(q)$ is responsible for non-adiabatic effects \cite{berczgama16B}, i.e. the dynamical coupling between the isotropic state and the anisotropic ones. Therefore a first approximation consists in neglecting $\hat{\mathrm{A}}(q)$ in the equations (\ref{b9dynamical}). In this case the system becomes separable and it results  the following solutions. The Friedman equation reduces to
\begin{equation}
\dfrac{1}{4}\textrm{H} ^2+ \dfrac{\widetilde{W}(q)}{q^2} - \dfrac{E_{n_0}(q)}{q^2} = 0\,,
\end{equation} 
where $n_0$ is a fixed value of the quantum number $n$, while the state $ | \phi^{{\rm (anis)}} (t)  \rangle$ evolves as
\begin{equation}
 | \phi^{{\rm (anis)}} (t) \rangle = \exp \left(\ii \int_0^t {N}(\tau) E_{n_0}(q_\tau) {\rm d}\tau \right)  |e_{n_0}^{{\rm (anis)}}(q_0) \rangle \,.
\end{equation}
Only one quantum level $n_0$ of the anisotropic Hamiltonian is involved in the dynamics and the eigenenergy $E_{n_0}(q)$ follows adiabatically the change of $q(t)$ during evolution. This corresponds to the Born-Oppenheimer approximation in Molecular Quantum Physics. Thanks to the repulsive part $\propto~  q^{-2}$ of the potential $\widetilde{W}(q)$ and the limit $\lim_{q \to 0} q^2 \, E_{n_0}(q) = 0$, the repulsive effect is always dominant near $q=0$ and the classical singularity is cured. It is replaced by a quantum bounce (see Fig. \ref{figBIXadiab}).

\subsubsection{Nonadiabatic (vibronic) approximation \cite{berczgama16B}}
\begin{figure}[htb]
\center
\includegraphics[width=0.4\textwidth]{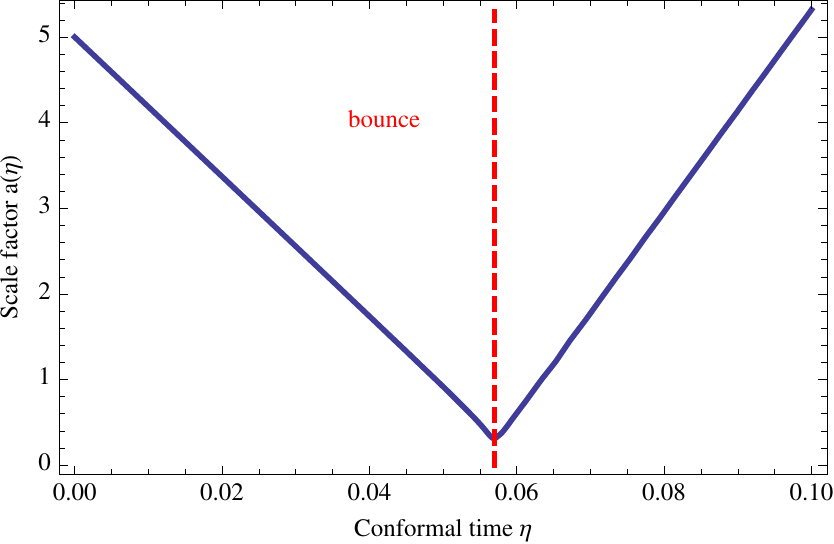}
\hspace{1cm}
\includegraphics[width=0.4\textwidth]{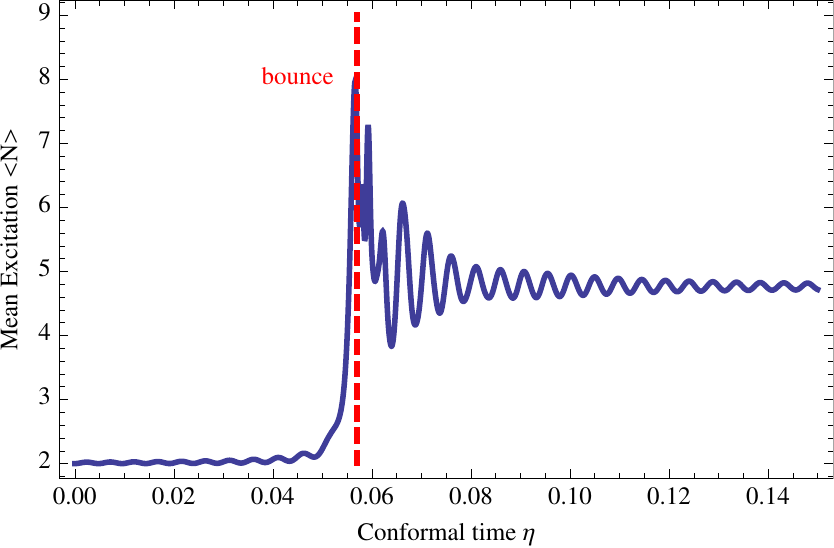} 
\caption{Non-adiabatic (vibronic-like) approximation: on the left panel plot of the scale factor $a=q^{2/3}$ as a function of time during a bounce, on the right panel plot of the degree of excitations through the same bounce. (Source: \cite{berczgama16B})}
\label{figBIXvib}
\end{figure}

 If we take into account the coupling due to $\hat{\mathrm{A}}(q)$, possible excitations and decays of anisotropic states are allowed during evolution. The system cannot be solved analytically anymore and only numerical simulations are available. We assume that the state  $ | \phi^{{\rm (anis)}} (t) \rangle$ is a finite sum $ | \phi^{{\rm (anis)}} (t) \rangle = \sum_n c_n(t) |e_n^{{\rm (anis)}}(q_0) \rangle$,  the functions $c_n(t)$ being numerically calculated. This corresponds to the vibronic framework in Molecular Quantum Physics. This procedure is presented in \cite{berczgama16B} where the harmonic approximation to the potential $V(\beta_+,\beta_-)$ is used. This approximation allows to obtain analytical formula for the eigenenergies $E_n(q)$, the eigenvectors $| e_n^{{\rm (anis)}}(q) \rangle$ and the operator $\hat{\mathrm{A}}(q)$. We show in \cite{berczgama16B} that even if the adiabatic approximation is broken (excitations and decays of anisotropy levels are allowed), the classical singularity is still replaced with a quantum bounce (see Fig. \ref{figBIXvib}).

The numerical simulations made for this  model strongly suggest  that the excitations of anisotropy due to nonadiabatic effects can be very large. Unfortunately, this specific numerical framework is not really suitable for the study of large excitations. Indeed,  the latter involve a large number of components in the sum $ | \phi^{{\rm (anis)}} (t) \rangle = \sum_n c_n(t) |e_n^{{\rm (anis)}}(q_0) \rangle$. Furthermore, the detailed knowledge of each component $c_n(t)$ becomes useless in the case of a large number of components. Therefore we need to develop another kind of approximation to study the domain of large excitations. This point deserves our full attention since it is highly related to the existence of an inflationary phase just after the bounce.\\


 \subsection{Nonadiabatic bounce and inflationary phase \cite{berczgama16A}}
\begin{figure}[htb]
\center
\includegraphics[width=0.6\textwidth]{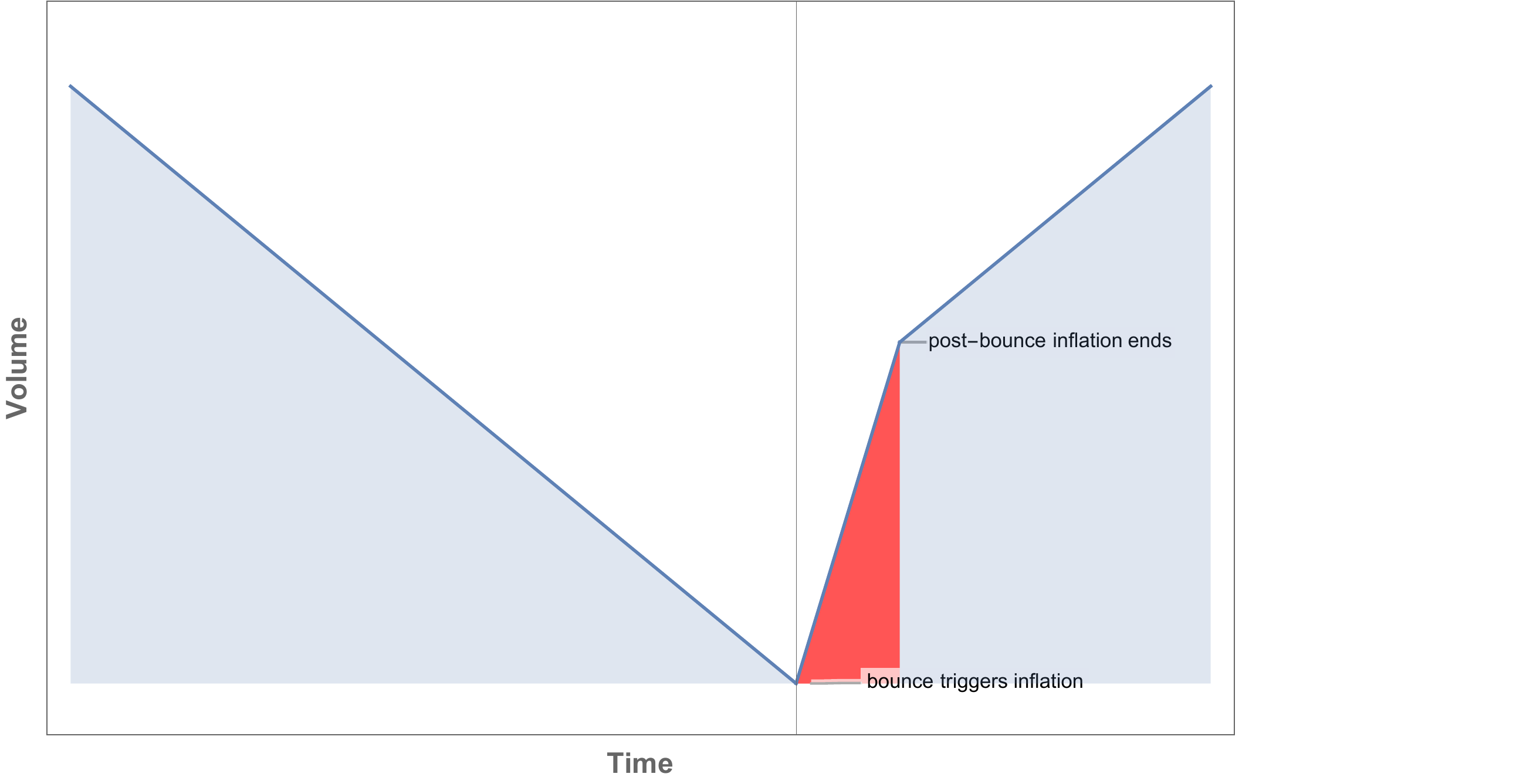}
\caption{The universe filled with radiation starts in adiabatic contraction with anisotropy in an eigenstate. Then, because of the repulsive potential, the bounce occurs and the anisotropy gets amplified. The produced anisotropy sources an inflationary phase, $\ddot{a} >0$, occurring just after the bounce. Later on, the anisotropy vanishes as $a^{-6}$, and the radiation again dominates the dynamics.}
\label{fignonadiab}
\end{figure}

 In  \cite{berczgama16A} we have proceeded with a first attempt to go beyond the adiabatic model in the case of large excitations. In other words, we let the energy levels of anisotropy vary with time as was indicated by our  study of the vibronic regime. This happens in response to the bounce, i.e. a sudden and significant change to the isotropic geometry described by $q$ and $p$. In principle the produced energy of anisotropy would gravitate and then influence the evolution of $q$ and $p$. However, we neglect this back-effect and solve the dynamics of $q$ and $p$ by keeping the energy level $N_i$ of anisotropy in $E_{N_i}(q)$ fixed, where $N_i$ is the initial number of anisotropic quanta. This framework allows us to address in a completely analytical manner many interesting questions. Three of them are of crucial importance: 
  \begin{enumerate}
  \item[(i)] What is the regime of validity of the adiabatic approximation? 
  \item[(ii)] What are the precise factors on which the excitation (or, decay) of anisotropy depends?
  \item[(iii)] What is the amount of anisotropic energy that can be produced in a violent bouncing cosmological scenario?
\end{enumerate} 
Our main finding is a sort of phase transition in the behavior of the anisotropic distortions. Once a critical value describing stiffness of the bounce is reached, the adiabatic (Born-Oppenheimer) approximation breaks down and a highly nonlinear excitation of anisotropic eigenstates takes place throughout the bounce. We considered a scenario in which the universe is isotropically and smoothly contracting in a FRW-like quantum state. The application of our result to this simple model of the universe shows that there occurs a large production of anisotropy at the bounce, which in turn leads to some sort of a sustained superexpansion phase similar to the one of the standard inflationary models (Fig. \ref{fignonadiab}).
More investigations are needed to study this inflationary phase. A semi-classical framework for both isotropic and anisotropic degrees of freedom seems to be the right way to take into account large excitations. However if we want to take into account at the same time the creation of anisotropy and the back-action on isotropy we need: 
\begin{enumerate}
  \item[(a)] to describe properly the quantum entanglement of degrees of freedom (operator $U(Q,q_0)$ in \eqref{coupledstate}) before building the semi-classical Hamiltonian, 
  \item[(b)] to avoid the harmonic approximation of the Bianchi IX potential which is broken for high excitations close to the bounce.
\end{enumerate}
This work is in progress.

\subsubsection{A first attempt to obtain a complete semi-classical framework}
In what follows we construct a complete semi-classical framework without resorting to any approximation to the anisotropy potential. For the half-plane $(q,p)$, we combine affine coherent state quantization based on a family of fiducial vectors labeled by $\mu$ with affine semi-classical portrait based on a family of affine coherent states built from a fiducial vector labeled by $\nu$ (the precise definitions of the fiducial vectors can be found in the appendices of \cite{grav_bounce}). For the planes $(\beta_{\pm},p_{\pm})$ we combine WHCIQ with Weyl-Heisenberg semi-classical portrait as explained in Sec. \ref{quantmod}.  

For the isotropic variables we find,
\begin{align}
\widecheck{(p^2)}=p^2+\frac{K(\mu,\nu)}{q^2},~\widecheck{(q^{\alpha})}=Q_{\alpha}(\mu,\nu)q^{\alpha},~K(\mu,\nu)=e^{\frac{3}{2\mu}}\left(\frac{\mu+\nu}{2}+\frac{1}{4}\right),~Q_{\alpha}(\mu,\nu)=e^{\frac{\alpha(\alpha-1)}{4\mu}}e^{\frac{\alpha(\alpha-1)}{4\nu}}.
\end{align}
For the anisotropic variables $(\beta_\pm, p_\pm)$ we pick the Gaussian weight function $\Pi$ with width $\sigma=\tau:=\frac{1}{2}\sigma_\pm$. The semi-classical portrait of the complete Hamiltonian constraint reads,
\begin{align}\label{vacC}
\widecheck{\mathrm{C}}=\frac{9}{4}\left(p^2+\frac{K(\mu,\nu)}{q^2}\right)-Q_{-2}(\mu,\nu)\frac{p_{\pm}^2+\frac{8}{\sigma_{\pm}^2}}{q^2}-36Q_{\frac{2}{3}}(\mu,\nu)q^{\frac{2}{3}}[\widecheck{V}(\bsb)-1]\, .
\end{align}
We note that the term $\propto \, q^{-2}$ is positive (i.e., produces a repulsion)  if and only if
\begin{align}
\frac{9}{4}K(\mu,\nu)>\sum_{\pm}Q_{-2}(\mu,\nu)\frac{8}{\sigma_{\pm}^2}.
\end{align}
The Hamilton equations read:
\begin{align}\label{fullsemi}
\begin{split}
\dot{q}&=\frac{9}{2}p,~\dot{p}=\frac{9}{2}\frac{K(\mu,\nu)}{q^3}-2Q_{-2}(\mu,\nu)\frac{p_{\pm}^2+\frac{8}{\sigma_{\pm}^2}}{q^3}+24Q_{\frac{2}{3}}(\mu,\nu)q^{-\frac{1}{3}}[\widecheck{V}(\bsb)-1],\\
\dot{\beta}_{\pm}&=-2Q_{-2}(\mu,\nu)\frac{p_{\pm}}{q^2},~\dot{p}_{\pm}=36Q_{\frac{2}{3}}(\mu,\nu)q^{\frac{2}{3}}\partial_{\pm}\widecheck{V}(\bsb),
\end{split}
\end{align}
where $\partial_{\pm}:=\partial_{\beta_{\pm}}$.
The vacuum Hamiltonian constraint (\ref{vacC}) may be supplemented with the radiation term $Rq^{-2/3}$.

The special case $\beta_{\pm}=0=p_{\pm}$ leads to a semi-classical model of the closed Friedmann universe, a special subclass of the Bianchi Type IX universes, given by the isotropic Hamiltonian constraint,
\begin{align}\label{isotropic}
\widecheck{\mathrm{C}}^{\mathrm{iso}}=\frac{9}{4}\left(p^2+\frac{K(\mu,\nu)-\frac{4}{9}Q_{-2}(\mu,\nu)\frac{8}{\sigma_{\pm}^2}}{q^2}\right)+36Q_{\frac{2}{3}}(\mu,\nu)q^{\frac{2}{3}}-\frac{R}{q^{2/3}}\, .
\end{align}

\begin{center}
\begin{figure}[h!]
\begin{tabular}{cc}
\includegraphics[width=0.28\textwidth]{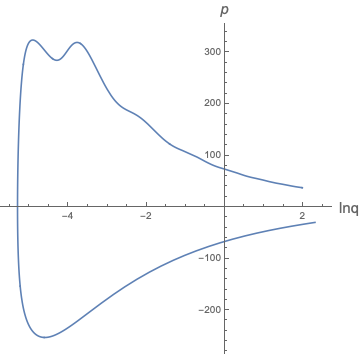}
\hspace{1cm}
\includegraphics[width=0.28\textwidth]{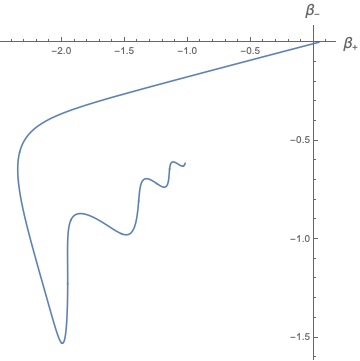}
\hspace{1cm}
\includegraphics[width=0.28\textwidth]{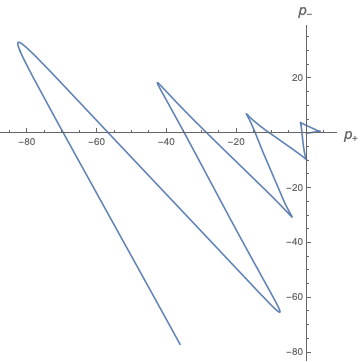}
\end{tabular}
\caption{A solution to the Hamilton eqs (\ref{fullsemi}) with the initial data: $\beta_+(0)=0.04$, $\beta_-(0)=0$, $p_+(0)=0$, $p_-(0)=0$. The model is fixed by setting $R=10^4$, $\sigma_{\pm}=100$ and $\mu=\nu=10$.} 
\label{sim2b}
\end{figure}
\end{center}

In Fig. (\ref{sim2b}) we plot an example of the solution to the complete semiclassical framework introduced above. In future works, we will proceed with a detailed analysis of different kinds of semi-classical models and their dynamical properties.

\section{Outlook}
 \label{perspec}
 
 The results described in the present review suggest that the construction of a minimal and realistic cosmological scenario able to account for the relevant cosmological data without fine tuning nor ad hoc assumptions is made possible with the utilization of the perturbed quantum mixmaster model. We have introduced important theoretical tools for deriving and solving quantum dynamics. With them we derived and studied the quantum and semi-classical dynamics of mixmaster. The reported herein findings encourage us to view the quantum mixmaster as a promising model of the primordial universe, in which the singularity is avoided in a way that creates a new and robust physical mechanism for generation of the primordial structure. This scenario requires only a minimal number of assumptions. Although the model shares some underpinning with alternative theories, the use of the quantum mixmaster bounce brings its own {\it qualitatively} distinctive characteristics.
 
 The next natural step of our research program is to study the evolution of local structures inside such a universe. For this purpose the Hamiltonian formalism for linear perturbations around mixmaster must be first developed. Then the quantized dynamics of the mixmaster background and linear perturbations thereon should be thoroughly investigated. Our expectation is that the bounce is going to strongly imprint on the spectrum of cosmological perturbations. At present we may only speculate that  the quantum bounce may produce a state of perturbations which for a certain range of cosmological scales is independent to the initial conditions, or which carries  detectable imprints of the primordial anisotropic oscillations, or which depends on the scale of the curvature at the bounce, etc. The key property to verify is whether the quantum mixmaster bounce may produce predictions for the primordial density perturbations consistent with the CMB results. Furthermore, the proposed model might do even better and, for instance, account for the widely discussed but essentially unexplained features of the primordial power spectrum such as large scale power suppression or localized oscillations \cite{hazra_etal14}, or explain why a fraction of the primordial density perturbation appears direction-dependent \cite{durakovica_etal18}. On the other hand, and this was already pointed out in the literature on bouncing cosmology, we should expect to obtain predictions for the spectrum of primordial gravitational waves (i.e., tensor perturbations), which are different from inflationary predictions. 
 
 In our approach the Universe is assumed to be spatially compact. This property of our model combined with quantum mechanics can lead us to discover relations between global quantities such as the volume of the Universe and local quantities such as the primordial power spectra as they all could be interconnected due to the quantum bounce. In particular, the ratio of the observable volume to the entire volume could be related to the amplitude of the primordial density perturbations, which would naturally explain the approximate flatness of the observable Universe. 
 
The presented model displays some new and interesting qualities which are absent in the other models of the primordial universe. Nevertheless, a large amount of work needs still to be done in order to justify or reject the model. For instance, sufficiently accurate solutions or semi-classical approximations to the quantum dynamics need still to be found with numerical methods. Then the next big step of deriving dynamics of perturbations and solving them has to be made. The investigation is in progress.


\subsection*{Acknowledgments}
The project is co-financed by the Polish National Agency for Academic Exchange and PHC POLONIUM 2019 (Project 42657QJ). \\ H. B. and J.-P. G. are grateful to National Centre for Nuclear Research, Warsaw (Poland) for hospitality during the preparation of the present work.

\section*{Abbreviations  used in this manuscript}

\begin{tabular}{@{}ll}
PGWs & Primordial gravitational waves\\
CMB & Cosmic microwave background\\
FRW & Friedman-Robertson-Walker\\ 
BKL &  Belinskii, Khalatnikov and Lifshitz\\
CS & Coherent states\\
ACS & Affine coherent states\\
WH & Weyl-Heisenberg\\
UIR & Unitary Irreducible representation\\
IR/UV & Infra-red/Ultra-violet
\end{tabular}

\appendix
\section{Toda approximation}\label{AppT}
The transformation to the new variables $q_1,~q_2,~q_3$ of Eq. (\ref{newq}) can be consistently extended by assuming an extra variable, say $\beta_z$, which is absent in the potential,
\begin{align}
\begin{bmatrix}
 \beta_+  \\
 \beta_-  \\
 \beta_z
 \end{bmatrix}
 =
\begin{bmatrix}
 \frac{1}{8} & 0 & - \frac{1}{8} \\
 \frac{1}{8\sqrt{3}} & -\frac{1}{4\sqrt{3}} & \frac{1}{8\sqrt{3}} \\
 a & b & c
 \end{bmatrix}
 \,
 \begin{bmatrix}
 q_1  \\
 q_2  \\
 q_3
 \end{bmatrix},
\end{align}
where $a,~b,~c$ are such that the transformation is invertible. Then we readily obtain the relation between conjugate momenta,
\begin{align}
\begin{bmatrix}
 p_1  \\
 p_2  \\
 p_3
 \end{bmatrix}
 =
\begin{bmatrix}
\frac{1}{8} &  \frac{1}{8\sqrt{3}} & a \\
 0 & -\frac{1}{4\sqrt{3}} & b \\
 - \frac{1}{8} & \frac{1}{8\sqrt{3}} & c
 \end{bmatrix}
 \,
 \begin{bmatrix}
 p_+  \\
 p_-  \\
 p_z
 \end{bmatrix},
\end{align}
and the Hamiltonian,
\begin{align}\label{focus2}
\mathrm{H}=32(p_1^2+p_2^2+p_3^2)+V_T+V_p+1,
\end{align}
which yields the anisotropic Hamiltonian (\ref{focus}) for the vanishing total momentum $p_1+p_2+p_3=0$ (or, $p_z=0$). The rescaling  $q_i\rightarrow \lambda q_i$, $p_i\rightarrow \lambda^{-1} p_i$ and $t\rightarrow 3e^{-\lambda} t$ such that $3e^{-\lambda}=\lambda^{2}$ brings the Hamiltonian (\ref{focus2}) to the following form (up to an irrelevant constant):
\begin{align}
\mathrm{H}=\frac{1}{2}(p_1^2+p_2^2+p_3^2)+e^{q_1-q_2}+e^{q_2-q_3}+e^{q_3-q_1}+3e^{-\frac{1}{2}\lambda}V_p.
\end{align}
The above Hamiltonian describes the periodic 3-particle Toda system \cite{berry76} plus another 3-particle Toda potential $3e^{-\frac{1}{2}\lambda}V_p$. 

\begin{figure}[!ht] 
\begin{center}
\includegraphics[scale=0.15]{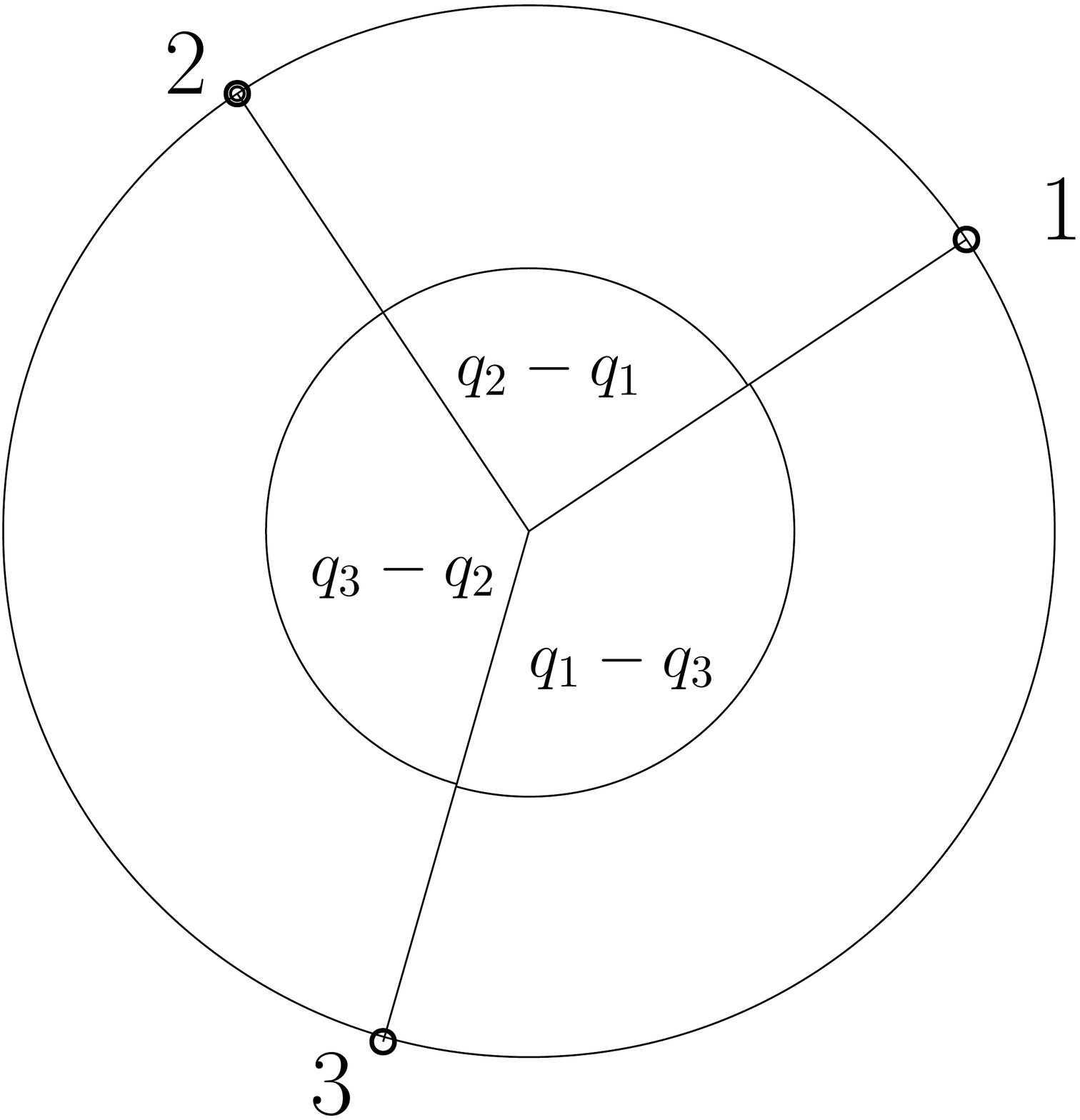} 
\end{center}
\caption{Model for the periodic 3-particle Toda lattice. The particles on the lattice  interact with their left and right  neighbour on the circle  via the exponential potential.}\label{3Toda}
\end{figure}

\section{Coefficients due to the fiducial vector of ACS}
\label{appendA}
\unskip
The vector $\psi$ of \eqref{acsdef} sets a family of coherent states in the Hilbert space of the quantum model. To obtain simple expressions for the constants, we can choose the following function of rapid decrease on $\mathbb{R}^+$, 
\begin{equation}
\label{psinu}
\psi_\nu(x) = \left( \frac{\nu}{\pi} \right)^{1/4} \frac{1}{\sqrt{x}} \exp \left[-\frac{\nu}{2} \left(\ln x - \frac{3}{4 \nu} \right)^2  \right] \quad \text{with} \quad \nu >0
\end{equation}
The above function is actually the square root of a Gaussian distribution on the real line with variable $y=\ln x$, centered at $3/4\nu$, and with variance $1/\nu$. With this function we obtain these constants as elementary functions of the free parameter $\nu$
\begin{equation}
\label{Ki}
\mathrm{k}_1 =  \frac{2 \nu+1}{4}, \quad \mathrm{k}_2=\exp \left[ \frac{3}{2\nu}\right], \quad \mathrm{k}_3
=\exp \left[- \frac{1}{18\nu}\right] \,.
\end{equation}
They are also given in Table \ref{ConstK} together with  5 other similar constants whose appearance through various expressions in the article results from our ACS approach.

\begin{table}[t]
  \centering
\begin{tabular}{|c|c|}
    \hline
$\mathfrak{K}_1(\nu)$   & $ \dfrac{2\nu + 1}{4}$   \\
   \hline
  $\mathfrak{K}_2(\nu)$   & $\exp \left[ \dfrac{3}{2\nu}\right]$   \\
   \hline
$\mathfrak{K}_3(\nu)$   & $\exp \left[ -\dfrac{1}{18\nu}\right]$   \\
   \hline
$\mathfrak{K}_4(\nu)$   & $ \left(\nu+ \dfrac{1}{4} \right) \exp \left[ \dfrac{3}{2\nu} \right]$   \\
   \hline
$\mathfrak{K}_5(\nu)$   & $\exp \left[- \dfrac{1}{9 \nu}\right]$   \\
   \hline
$\mathfrak{K}_6(\nu)$   & $\sqrt{2}\,
\exp \left[ \dfrac{1}{\nu}\right]$   \\
   \hline
   $\mathfrak{K}_7(\nu)$   & $\exp \left[ \dfrac{1}{2\nu}\right]$   \\
   \hline
$\mathfrak{K}_8(\nu)$   & $ \exp \left[ \dfrac{3}{2\nu}\right]$   \\
   \hline
\end{tabular}
  \caption{Constants $\mathfrak{K}_i\equiv \mathfrak{K}_i(\nu)$, $i=1,2, \dotsc, 8$ as functions of the free parameter $\nu$ appearing in the fiducial vector \eqref{psinu}}\label{ConstK}
\end{table}
\newpage



\end{document}